\documentclass[aps,prd,preprint,tightenlines,superscriptaddress,
amsfonts,amssymb,amsmath,byrevtex,showpacs,nofootinbib]{revtex4-2}
\usepackage[polish,english]{babel}
\usepackage[utf8]{inputenc}
\usepackage[T1]{fontenc}
\usepackage{ulem}
\usepackage{latexsym}
\usepackage{graphicx}
\usepackage{geometry}
\usepackage{subfig}
\usepackage{color}

\graphicspath{ {./}{./rys/} } % wzgledna sciezki do grafiki
\usepackage{xcolor}
\usepackage{pdfsync}
\usepackage{fancyhdr}
\usepackage{makeidx}
\usepackage[breaklinks,colorlinks,backref]{hyperref}
\hypersetup{
    colorlinks, %set true if you want colored links
    linktoc=all, %set to all if you want both sections and subsections linked
    linkcolor=red,  %choose some color if you want links to stand out
  citecolor=hyptxt,
  urlcolor=blue}
  \hypersetup{
  citebordercolor=,
  filebordercolor=green,
  linkbordercolor=blue
}
\definecolor{hyptxt}{rgb}{0.7, 0.4, 0.9}
\usepackage{bibtopic}
\usepackage{amsmath}
\newcommand{\RNumb}{\mathbb{R}}

\newcommand{\UnitOp}{\hat{1\kern-4.75pt 1}} % operator unit with hat
\newcommand{\MatUnit}{1\kern-3pt 1} % matrix unit
\newcommand{\Group}[1]{\textrm{#1}} % group assignment

\newcommand{\HW}[1]{\mathcal{HW}(#1)} % Heisenberg-Weyl group
\newcommand{\Bra}[1]{\langle #1 \vert} % bra
\newcommand{\Ket}[1]{\vert #1 \rangle} % ket
\newcommand{\BraKet}[2]{\langle #1 \vert #2 \rangle} % bra(c)ket
\newcommand{\Norm}[1]{\|\kern.3ex#1\kern.3ex \|} % norm macro
\newcommand{\EOp}{\mathsf{E}\kern-1pt\llap{$\vert$}}   % PEv E operator
\newcommand{\WOp}{\hat{\mathsf{W}\kern-1pt\llap{$-$}}} % PEv W generator
\newcommand{\FOp}{\mathsf{F}\kern-1pt\llap{$\vert$}}   % QEv F operator
\newcommand{\StateSpace}[1]{\mathcal #1} % state space
\newcommand{\Komentarz}[1]{} % comment

\usepackage{accents}

\begin{document}

\title{Semiclassical causal geodesics: Minkowski spacetime case}

\author{Adam Cie\'{s}lik} \email{adam.cieslik@univie.ac.at}
\affiliation{Institute of Theoretical Physics, Jagiellonian University in
  Krak\'{o}w, {\L}ojasiewicza 11, 30-348 Krak\'{o}w, Poland}
\affiliation{Faculty of Physics, University of Vienna, W\"{a}hringer Stra{\ss}e 17, 1090 Wien, Austria}

\author{Andrzej G\'{o}\'{z}d\'{z}}
\email{andrzej.gozdz@umcs.lublin.pl}
\affiliation{Institute of Physics, Maria Curie-Sk{\l}odowska
University, pl.  Marii Curie-Sk{\l}odowskiej 1, 20-031 Lublin, Poland}

\author{Patryk Mach} \email{patryk.mach@uj.edu.pl} \affiliation{Institute of
  Theoretical Physics, Jagiellonian University in Krak\'{o}w,
  {\L}ojasiewicza 11, 30-348 Krak\'{o}w, Poland}

\author{Aleksandra P\c{e}drak} \email{aleksandra.pedrak@ncbj.gov.pl}
\affiliation{Department of Fundamental Research, National Centre for Nuclear
  Research, Pasteura 7, 02-093 Warszawa, Poland}

\author{W{\l}odzimierz Piechocki} \email{wlodzimierz.piechocki@ncbj.gov.pl}
\affiliation{Department of Fundamental Research, National Centre for Nuclear
  Research, Pasteura 7, 02-093 Warszawa, Poland}

%%%%%%%%%%%%%%%%%%%%%%%%%%%%%%%%%%%%%%%%%%%%%%%%%%%%%%%%%%%%%%%%%%%%%%%%
\date{\today}
%%%%%%%%%%%%%%%%%%%%%%%%%%%%%%%%%%%%%%%%%%%%%%%%%%%%%%%%%%%%%%%%%%%%%%%%

\begin{abstract}
We use an integral quantization model based on the Heisenberg-Weyl group to
describe the motion of a spinless particle in the Minkowski
background spacetime.  This work is a sequel to a previous paper, devoted to
mathematical aspects of our model: construction of the space of coherent states
and properties of elementary observables.  We compute transition amplitudes
corresponding to a free motion of a particle between two coherent states. These
amplitudes are then used to model quantum random walks of free relativistic
particles. Our quantization scheme allows us to recover interference patterns
occurring in a standard double-slit experiment, known from the classical
approach.  This result is obtained by modeling the slits in terms of eigenstates
of the position operator and computing transition amplitudes between position
and coherent states.  We design our model in a way which allows for a future
generalization to a semi-classical quantization of the geodesic motion in curved
spacetimes.
\end{abstract}

%%%%%%%%%%%%%%%%%%%%%%%%%%%%%%%%%%%%%%%%%%%%%%%%%%%%%%%%%%%%%%%%%%%%%%%%%%%%
%\pacs{XXX}
%%%%%%%%%%%%%%%%%%%%%%%%%%%%%%%%%%%%%%%%%%%%%%%%%%%%%%%%%%%%%%%%%%%%%%%%%%%%%

\maketitle

\tableofcontents

%%%%%%%%%%%%%%%%%%%%%%%%%%%%%%%%%
\section{Introduction}\label{Int}
%%%%%%%%%%%%%%%%%%%%%%%%%%%%%%%%%

% General relativity (GR) proves to be successful in describing most of known
% classical gravitational systems, especially with low curvature of
% spacetime. Quantum theory (QT) describing, for instance, the world of
% elementary particles in the flat spacetime is successful as well. It is
% believed that near gravitational singularities, where GR breaks down, quantum
% effects should be taken into account and both theories GR and QT should be
% unified into a theory of quantum gravity (QG). However, this idea turns out to
% be enormously difficult in realization.

The struggle for the construction of a generally accepted theory of quantum
gravity lasts for more than fifty years.  The hope that string theory (see,
e.g., \cite{Green}), loop quantum gravity (see, e.g., \cite{Gam,Thi,Rov}),
causal dynamical triangulations \cite{cdt21,cdt12}, integral quantization (to be
applied in the present paper), or other theory could play that role has not been
fulfilled yet. The main reason seems to be related to the lack of experimental
data concerning extremal gravitational fields, which could be used to constrain
those theories.

On the other hand, recent observational advances---the detection of
gravitational waves or observations of shadows of supermassive black holes
$M87^*$ and $Sgr A^*$ by the Event Horizon Telescope (EHT)
\cite{M87,Sgr}---might provide new data in a range that could become relevant
for semiclassical models. Here by a semiclassical model we understand a
description in which only a part of the physical system is described by a
quantum theory, while the remaining part is treated classically.  In the context
of gravitational physics, the best-known example of a semiclassical approach is
provided by the quantum field theory in curved spacetime \cite{birell,Robert},
in which the gravitational field is treated classically. A converse situation is
also possible---see, e.g., \cite{George1,George2} for an analysis of quantum
fluctuations of the gravitational field affecting geodesic motion of test
particles.

Our work, presented in this paper, has been motivated by an attempt to quantize
geodesic motion of test particles in a fixed (classical) spacetime, employing an
integral quantization based on coherent states. If successful, such a program
could, in principle, lead to observational consequences. The gravitational field
in the vicinity of black holes is sufficiently strong to trap photons in a
quasi-periodic motion around them. The set of all such trapped photon orbits has
been found to be related to the shadows of the black holes
\cite{Per5,Gre5}. Properly quantized trapped photon regions might reproduce
observed shadows better than the corresponding classical theory.

On the other hand, the geodesic description of motion should be understood as
an approximation even for test matter in the classical regime. This is especially
clear for the electromagnetic radiation, which can also be modelled by directly
solving Maxwell equations on a fixed background spacetime. It is known that the
geometrical optics approximation can break down for wavelengths comparable to
a characteristic length scale associated with the spacetime curvature, leading
to backscattering of electromagnetic waves on the spacetime geometry \cite{malec1,malec2,malec3}.

For many classical spacetimes, including Schwarzschild and Kerr solutions,
causal geodesics are well known (see, e.g., \cite{hagihara_theory_1930,Carter1968},
classic textbooks \cite{chandrasekhar_mathematical_1983,Neill1995}
or recent accounts in \cite{CM2022,Kerr23,Bakun25} and references therein). However,
promoting the classical description of a geodesic motion to a semiclassical
framework is challenging, and a suitable formalism should first be
established. It seems that a reasonable strategy---adopted in this
paper---would be to start with finding the quantum approach for the case of a
test particle geodesic motion in the Minkowski spacetime. Generalizations to
curved spacetimes will be presented elsewhere.

Basic elements of our construction adapted to the flat Minkowski spacetime were
recently discussed in \cite{IQmethod}.  The configuration space of coherent
states was obtained from an action of the four-dimensional Heisenberg-Weyl
group.  A suitable version of the Positive Operator Valued Measure (POVM)
formalism was applied to elementary observables---components of positions and
four-momenta---and to the quantum counterpart of a classical Hamiltonian
associated with geodesic equations.

In this work, we apply the formalism introduced in \cite{IQmethod} to compute
probability amplitudes corresponding to transitions of a free particle from one
coherent state to another. This construction allows us to deal with appropriate
mass-shell conditions. As a test, we recover interference patterns occurring in
the standard double-slit experiment.

This paper is organized as follows. After introducing the notation and preliminary
notions in Sec.\ II, we recall, in Sec.\ III, some aspects of the integral quantization method
based on the Heisenberg-Weyl group. They include the unitary
representation of that group, a construction of coherent states, and a
mapping of classical observables into quantum operators. In Section IV a spectrum of a Hamiltonian associated with the geodesic motion is computed. In Section V we define the
transition probability of a test particle. Section VI is devoted to an
application of our quantum formalism: an evolution of a test particle, finding
stochastic trajectories, and computing the interference patterns in a simple
double-slit experiment. We conclude in Sec. VII.  Appendix A concerns the
overlap of quantum points (coherent states) in our configuration space. Appendix
B presents a transition probability of a massless test particle.

We used Wolfram Mathematica \cite{Wolfram} to perform our numerical calculations.
A demonstration notebook containing our numerical code and data will be publicly available in \cite{Notebook}.

%%%%%%%%%%%%%%%%%%%%%%%%%%%%%%%%%%%%%%%%%%%%%%%%%%%%%%%%%
\section{Preliminaries}\label{Pre}
%%%%%%%%%%%%%%%%%%%%%%%%%%%%%%%%%%%5%%%%%%%%%%%%%%%%%%%%%

Let $(\mathcal M, g)$ denote a spacetime manifold. In general relativity, or
more generally on curved spacetimes, geodesic equations can be written in a
Hamiltonian form as
\begin{equation}
\label{gHamEqs}
\frac{d x^\mu}{d \tilde s} = \frac{\partial H}{\partial p_\mu}, \quad  \frac{d p_\nu}{d \tilde s} =
- \frac{\partial H}{\partial x^\nu},
\end{equation}
where the Hamiltonian $H$ has the form
\begin{equation}
\label{gHam}
H(x,p) = \frac{1}{2} g^{\mu\nu}(x) p_{\mu} p_{\nu}.
\end{equation}
Here geodesics are understood as curves
$\mathbb R \supseteq I \ni \tilde s \to x(\tilde s) \in \mathcal M$.  The
four-momentum components $p_\mu$ are defined by $p^\mu = dx^\mu /d \tilde
s$. For convenience, we assume a normalization convention with
$H = - \frac{1}{2} \, m^2$, where $m \ge 0$ denotes the particle rest mass.
Timelike and null geodesics are
characterized by $m > 0$ and $m = 0$, respectively. This condition and the
definition of the four-momenta imply that $\tilde s = s/m$, where $s$ denotes
the proper time.

In this paper, we restrict ourselves to the four-dimensional flat Minkowski
spacetime. We assume a convention with the metric signature $(-,+,+,+)$, so that
in an orthonormal frame one has
$g^{\mu\nu} p_\mu p_\nu = - p_0^2 + p_1^2 + p_2^2 + p_3^2$. In this case,
solutions of geodesic equations (\ref{gHamEqs}) correspond to straight lines
given by
\begin{equation}
\label{HamSol}
p_\mu = \mathrm{const}, \quad x^\mu = p^\mu \tilde s, \quad \mu = 0,1,2,3.
\end{equation}
For timelike geodesics, one can write $x^\mu = p^\mu s/m$.

The above Hamiltonian formalism gives rise to a notion of the phase space,
identified naturally with the cotangent bundle defined as
\begin{equation}
T^\star \mathcal M = \{ (x,p) \colon x \in \mathcal M, \, p \in T_x^\star \mathcal M \},
\end{equation}
where $T_x^\star \mathcal M$ denotes the cotangent space at $x \in \mathcal
M$. In the case of the Minkowski spacetime we have
$T^\star \mathcal M \cong \RNumb^4 \times \RNumb^4$. Throughout this paper, we
treat the conjugate variables $p_\mu$ and $x^\nu$ as independent.

In numerical calculations we apply the Planck units by setting $c=1=G=\hbar$,
where $c$ is the speed of light, $G$ is the gravitational constant, and $\hbar$
denotes the Planck constant. This choice of units renders the set
$\{x^\mu, p_\mu, m \}$ dimensionless. The conversion of formulas from Planck
to standard units can be done by using, e.g., App.\ F of
\cite{Wald}.

%%%%%%%%%%%%%%%%%%%%%%%%%%%%%%%%%%%%%%%%%%%%%%%
\section{Integral quantization}\label{Integral}
%%%%%%%%%%%%%%%%%%%%%%%%%%%%%%%%%%%%%%%%%%%%%%%

In this paper we use the so-called integral quantization method. Sample
applications of the integral quantization in astrophysical and cosmological
contexts can be found in \cite{ast1,ast2,ast3,ast4} and
\cite{cos1,cos2,cos3,met}, respectively. The mathematical background and many
details of our scheme are presented in \cite{IQmethod}.  For similar
approaches to the integral quantization we recommend the papers
\cite{hyb1,hyb2,hyb3,hyb4,hyb5}.

A general idea of the integral quantization requires the existence of a
one-to-one transformation of the space of elementary variables of a physical
system under consideration (extended configuration or phase space) onto a group
$\Group{G}$. The group $\Group{G}$ should have a unitary irreducible
representation (UIR) in a carrier Hilbert space $\StateSpace{K}$, which allows
for a construction of the space of coherent states in $\StateSpace{K}$.

In this paper, we choose the Heisenberg-Weyl group $\HW{4}$ as the group
$\Group{G}$, since it can be identified with the classical phase space
$T^\star \mathcal{M}$. To some extent, the integral quantization based on the
Heisenberg-Weyl group can be treated as a generalization of the canonical
quantization.  In papers \cite{ast1,ast2,ast3,ast4,cos1,cos2,cos3,met} we have
used the affine group.

The group $\HW{4}$ is known to have a UIR in the Hilbert space
$L^2(\RNumb^4,d^{4}\xi) =: \StateSpace{K}$, where
$d^{4}\xi:= d\xi^0\,d\xi^1\dots d\xi^{3}$.  This representation can be
defined as
\begin{equation}
\label{rep0}
\hat{\mathcal{U}}(\kappa;p,x)=\exp(i\kappa \UnitOp)\, \hat{\mathcal{U}}(p,x) \, ,
\end{equation}
where
\begin{equation}\label{rep1}
 \hat{\mathcal{U}}(p,x)\psi(\xi)=\exp\left(\frac{-ip_\mu x^\mu}{2\hbar}\right)
\exp\left(\frac{ip_\mu \xi^\mu}{\hbar}\right)\psi(\xi-x) \, ,
\end{equation}
$\UnitOp$ is the unit operator,  and the states $\Ket{\psi}$ are defined as
$\psi (\xi):= \BraKet{\xi}{\psi} \in \StateSpace{K} $.

The coherent states,  $\Ket{p,x} \in \mathcal{K} $, are defined by
\begin{equation}\label{rep2}
 \Ket{p,x}=\hat{\mathcal{U}}(p,x)\Ket{\Phi_0}, \quad \BraKet{\xi}{p,x}
=\hat{\mathcal{U}}(p,x) \BraKet{\xi}{\Phi_0}
=\hat{\mathcal{U}}(p,x)\Phi_0(\xi)\, ,
\end{equation}
where $\Phi_0(\xi) \colon \RNumb^4 \rightarrow \mathbb{C}$ is the so-called
fiducial vector $|\Phi_0 \rangle \in \mathcal{K}$, such that
$\langle \Phi_0 | \Phi_0 \rangle = 1$.  The fiducial vector can be understood as
a parameter of the quantization based on coherent states.  The $\xi$-independent factor
$\exp\left(\frac{-ip_\mu x^\mu}{2\hbar}\right)$ appearing in Eq.\ (\ref{rep1}) is in fact
an arbitrary global phase. It was chosen following a convention from \cite{Per}.
On the other hand, in Quantum Mechanics, measurable physical quantities remain independent
of global phases of quantum states. In addition, in our case, the resolution of unity and,
more generally, the integral quantization scheme are also independent of the choice of such phase factors.

Since the representation is irreducible, the operators
$| p,x \rangle \langle p,x | \colon \StateSpace{K} \rightarrow \StateSpace{K}$
satisfy
\begin{equation}
\label{5resolution}
(2\pi\hbar)^{-4}\int_{\RNumb^{8}}\, d\rho(p,x) \,\Ket{p,x} \Bra{p,x}
=\UnitOp \, ,
\end{equation}
where $d\rho(p,x):= dp_0\,dp_1\dots dp_{3}\,d x^0\, dx^1\dots dx^{3}$. Equation
(\ref{5resolution}) provides the resolution of the unity operator in
$\StateSpace{K}$.

The Heisenberg-Weyl quantization consists of assigning uniquely to each point of
the phase space $T^\star \mathcal{M}$ the projection operator
\begin{equation}\label{acs5}
  \RNumb^{8} \ni (p,x) \longrightarrow \Ket{p,x}\Bra{p,x}  \, .
\end{equation}
Equation \eqref{5resolution} can be used for mapping (quantization) of almost
any classical observable $f \colon \RNumb^{8} \rightarrow \RNumb$ onto an
operator $\hat{f} \colon \StateSpace{K} \rightarrow \StateSpace{K}$ as follows
\begin{equation}
\label{5mapping}
 f \longrightarrow \hat{f} :=
 (2\pi\hbar)^{-4}\int_{\RNumb^{8}}\, d\rho(p,x)
 |p,x\rangle f(p,x)\langle p,x| \, .
\end{equation}
The mapping \eqref{5mapping} defines a symmetric operator, and if the classical
observable $f(p,x)$ is either a bounded or an integrable function,
$f(p,x) \in L^1(\RNumb^{8},d\rho(p,x))$, the mapping \eqref{5mapping} defines a
self-adjoint operator. If $\hat{f}$ is not self-adjoint, the arising difficulty
can be solved, for instance, by making use of the theory of so-called positive
operator-valued measures.  We recommend our recent paper \cite{IQmethod} for
more details.

To fix the meaning of quantum numbers $p$ and $x$ labeling coherent states
\eqref{rep2}, we use the following consistency condition:
\begin{equation}
\label{CSConsistCon}
 \Bra{p,x} \hat{p}^\mu \Ket{p,x} =p^\mu, \quad
 \Bra{p,x} \hat{x}^\mu \Ket{p,x} =x^\mu \,,
\end{equation}
where $\mu=0,1,2,3$. Conditions \eqref{CSConsistCon} mean that $p_\mu$ and
$x^\mu$ represent expectation values of four-momenta and four-positions in the
coherent states.

%%%%%%%%%%%%%%%%%%%%%%%%%%%%%%%%%%%%%%%%%
\section{Spectrum of the geodesic Hamiltonian}
%%%%%%%%%%%%%%%%%%%%%%%%%%%%%%%%%%%%%%%%%

Equation (\ref{gHam}) of Sec.\ \ref{Pre} defines a Hamiltonian $H(p,x)$
corresponding to the equations of motion of a test particle---geodesic equations
(\ref{gHamEqs}). Due to Eq.\ \eqref{5mapping}, the corresponding quantum
Hamiltonian $\hat{H}$ has the form
\begin{equation}\label{Ham2}
  \hat{H} = (2\pi\hbar)^{-4} \int_{\RNumb^{8}}\, d\rho(p,x)
  \Ket{p,x}H(p,x)\Bra{p,x} \, .
\end{equation}
One can show \cite{IQmethod} that the functions defined as
\begin{equation}
\label{Ham3}
 \eta_p(\xi)=\BraKet{\xi}{\eta_p}
:= \left( \frac{1}{\sqrt{2\pi \hbar}} \right)^4 \exp \left( i\frac{p\, \xi}{\hbar} \right)
\, ,
\end{equation}
where $p\, \xi := p_{\mu} \xi^\mu$, with $\mu = 0,1,2,3$, are generalized
eigenstates of $\hat{H}$, defined by \eqref{Ham2}. The key element in this
reasoning is to make use of the orthogonal decomposition of the unity in the
carrier space $\mathcal{K}$ in terms of the generalized states \eqref{Ham3}:
\begin{equation}
\label{EtaUnity}
\int_{\RNumb^4} d^4p\, \Ket{\eta_p}\Bra{\eta_p}=\UnitOp \, .
\end{equation}
The validity of \eqref{EtaUnity} stems from the theory of Fourier transforms in
the context of distributions (see, e.g., \cite{Rob}).

One can show \cite{IQmethod} that the eigenvalue problem for the Hamiltonian
\eqref{Ham2} can be written as
\begin{eqnarray}
\label{HamAct5}
&&  \left(\int_{\RNumb^{4}}d^4 p\, |\tilde{\Phi}_0 (p)|^2\right)
 g^{\alpha\beta} k_{\alpha} k_{\beta}
+ \left( \int_{\RNumb^{4}}d^4 p\,p_\beta |\tilde{\Phi}_0 (p) |^2 \right)
2 g^{\alpha\beta} k_{\alpha}  \nonumber \\
&& + \int_{\RNumb^{4}} d^4 p\, g^{\alpha\beta} p_\alpha p_\beta
|\tilde{\Phi}_0 (p) |^2 \, =  - m^2\, ,
\end{eqnarray}
where $\tilde{\Phi}_0 (p)$ is the Fourier transform of some fiducial vector
${\Phi}_0 (\xi)$. Equation \eqref{HamAct5} can be simplified by a suitable
choice of ${\Phi}_0 (\xi)$.  A remarkably simple form of Eq.\ \eqref{HamAct5} is
obtained by selecting the fiducial vector in the form of the wave function
corresponding to the ground state of the 4D harmonic oscillator, i.e.,
\begin{equation}
\label{Ham17}
 \Phi_0 (\xi) =  \prod_{\mu=0}^3
 \left(\frac{\lambda_\mu}{\pi\hbar}\right)^{\frac{1}{4}}
\exp\left(-\frac{\lambda_\mu (\xi^\mu)^2}{2\hbar}\right)\,,
\end{equation}
with $\lambda_0 = 3 \lambda_1$ and $\lambda_1=\lambda_2=\lambda_3 >0$. In this
case, the functions \eqref{Ham3} are the eigenstates of the Hamiltonian
\eqref{Ham2}, provided that
\begin{equation}\label{finally}
 g^{\alpha\beta}p_\alpha p_\beta = - m^2 \,,
\end{equation}
in agreement with the relation satisfied by the classical momenta
\cite{IQmethod}.

The quantum Hamiltonian $\hat{H}$ has a continuous spectrum consisting of
eigenvalues $-\frac{1}{2}m^2$, each of which is infinitely many-fold degenerate.\\

%{\bf 
In principle, our integral quantization method can be applied to non-relativistic cases as well. For this purpose one can use, for instance, the reduction procedure proposed by Horwitz and Rotbart \cite{non-rel}, to derive the non-relativistic limit of relation (\ref{finally}). On a more fundamental level, we note that while up to this point our quantization scheme relies heavily on the affine
structure of the spacetime, little use is made of the assumed metric. The Minkowski metric is used to define
the Hamiltonian $H$, and it will be used to construct transition amplitudes respecting the mass-shell condition.
This leaves open a possibility to consider non-relativistic spacetimes as well. In this case, usual requirements of invariance
under spacetime translations and space rotations are compatible with the non-relativistic Hamiltonian of the form
\begin{equation}
\label{NonRelHam1}
H_\mathrm{NR}(p_0,\mathbf p)= A p_0 + \frac{1}{2} B \mathbf{p}^2,
\end{equation}
where $p_\mu = (p_0,\mathbf p)$ and $A$ and $B$ are constants\footnote{For a more detailed analysis of possible
kinamatical symmetries see, e.g., \cite{Bacry}.}. In the Minkowski spacetime, the Poincar\'{e} symmetry implies,
for a spinless free particle, the Hamiltonian of the form (\ref{gHam}). In both cases, solutions of Hamilton's
equations correspond to straight lines and constant momentum components $p_\mu$.

Our integral quantization of the non-relativistic Hamiltonian \eqref{NonRelHam1} leads to the Schr\"odinger type
operator $\hat{H}_\mathrm{NR} = A \hat{p}_0 +\frac{1}{2} B \sum_{l=1}^3\widehat{p_l^2}$ and
the Klein-Gordon type operator in case of (\ref{gHam}). One can directly show that both Hamiltonians have the
same eigenfunctions represented by the plain waves $\Ket{\eta_k}$, but their eigenvalues are different.

Further discussion of the non-relativistic case is beyond the scope of the present paper, but will be discussed elsewhere.
%}

%%%%%%%%%%%%%%%%%%%%%%%%%%%%%%%%%%%%%%%%%%%%%%%%%
\section{Transition probability of a test particle}
\label{sec:Transition_prob}
%%%%%%%%%%%%%%%%%%%%%%%%%%%%%%%%%%%%%%%%%%%%%%%%%

We will now apply our formalism to a computation of transition amplitudes
between coherent states, belonging to the Hilbert space $\mathcal{K}$.

We begin by introducing the mass layer  $\mathcal{J}_{m,\epsilon}$ of
thickness $\epsilon$, describing a test particle $m \geq 0$. It is defined as
\begin{equation}
\label{eq:MassLayer}
\mathcal{J}_{m,\epsilon}
:= \left\{ p  \colon -\sqrt{m^2+\mathbf{p}^2+\epsilon} \leq p_0
\leq -\sqrt{m^2+\mathbf{p}^2}, \,\, \mathbf{p} \in \RNumb^3  \right\} \,,
\end{equation}
and will be used as a subsidiary set allowing for a construction of well-defined
transition amplitudes. Taking the limit of $\epsilon \rightarrow 0$ leads to
the commonly used notion of the mass shell. Our construction of the layer is
based on one of the solutions to the equation
$g^{\alpha\beta}k_\alpha k_\beta = - m^2$, which is compatible with the choice
of the metric signature $(-,+,+,+)$ and the orthochronous part of the Lorentz
group \cite{SW}.

The operator projecting onto the mass layer \eqref{eq:MassLayer} is constructed
from the generalized eigenstates of the test particle Hamiltonian, defined in
the preceding section, as follows
\begin{equation}
\label{eq:ProjMassLayer}
P_{\mathcal{J}_{m,\epsilon}} :=
\int_{\RNumb^4} d^4 p\,
\Ket{\eta_p} \chi(p \in \mathcal{J}_{m,\epsilon}) \Bra{\eta_p} \, ,
\end{equation}
where $\chi (p \in Q)= 1$ iff the relationship $Q$ is satisfied and equals $0$
otherwise. The transition amplitude of the particle of mass $m \geq 0$ from a
state $\Ket{p'x'}$ to a state $\Ket{p''x''}$ is given by the following matrix
element of the projection operator:
\begin{equation}
\label{Tran2}
\mathcal{A}_{m,\epsilon}
:=  \Bra{p'',x''} P_{\mathcal{J}_{m,\epsilon}} \Ket{p',x'} \,.
\end{equation}

The important properties of the transition amplitudes are their transformation
features.  Every transformation $T(p,x):=(T_p(p,x),T_x(p,x))$ of the classical
configuration space $T^\star \mathcal M $ defines a transformation in the
carrier space $\StateSpace{K}$:
\begin{equation}
\label{TransConfSp}
\Ket{T;p,x} \equiv \hat{T}\Ket{p,x}
:= \exp(i \phi_T(p,x)) \Ket{T_p(p,x),T_x(p,x)} \,.
\end{equation}
For every operator $\hat{A}:\StateSpace{K} \to \StateSpace{K}$, the matrix
elements of $\hat{A}$ between the states \eqref{TransConfSp} can be written as
products of phase factors and matrix elements of $\hat{A}$ between coherent
states corresponding to transformed momenta and positions
\begin{eqnarray}
\label{MatElemTransCS}
&&\Bra{T;p',x'} \hat{A} \Ket{T;p,x}
= \exp\left( i [\phi_T(p,x)-\phi_T(p',x')] \right) \nonumber\\
&& \Bra{T_p(p',x'),T_x(p',x')}\hat{A}\Ket{T_p(p,x),T_x(p,x)} \,.
\end{eqnarray}
Note that the expectation values between transformed coherent states are
independent of the phase appearing in \eqref{MatElemTransCS}. If $\hat{A}$ is a transition operator like
$P_{\mathcal{J}_{m,\epsilon}}$, the modulus square of \eqref{MatElemTransCS}
representing the transition probability is also independent of this phase. On the
other hand, an interference of two or more amplitudes \eqref{MatElemTransCS} depends on such phases.

All operators $\hat{\mathcal{U}}(\kappa;p,x)$ defined by Eq.~\eqref{rep0}
are of the form \eqref{MatElemTransCS}
\begin{equation}
\label{UActCS}
\hat{\mathcal{U}}(\kappa';p',x')\Ket{p,x}
=\exp\left(i\left[\kappa'-\frac{px'-p'x}{2\hbar}\right]\right)
\Ket{p+p',x+x'}\,.
\end{equation}
Similarly, a representation of the Poincar\'e transformation
$w(a,\Lambda)(p,x)=(p',x'):=(\Lambda p, \Lambda x +a)$ in the carrier space
$\StateSpace{K}$ reads
\begin{equation}
\label{PoincareTransf}
\hat{\mathcal{G}}(a,\Lambda) \Ket{p,x}
:= \exp(i \phi_{\mathcal{G}(a,\Lambda)}(p,x)) \Ket{\Lambda p,\Lambda x+a} \,,
\end{equation}
where $\Lambda$ denotes the Lorentz transformation and $a$ is the
four--translation vector. The phase factor has to be calculated from the
multiplication law of the Poincar\'e group:
$\hat{\mathcal{G}}(a',\Lambda')\hat{\mathcal{G}}(a,\Lambda)=
\hat{\mathcal{G}}(a'+\Lambda'a,\Lambda'\Lambda)$. This implies the following
condition for the phase factors
\begin{equation}
\label{PoincarePhase}
\exp(i \phi_{\mathcal{G}(a,\Lambda)}(p,x))
\exp(i \phi_{\mathcal{G}(a',\Lambda')}
(\Lambda p,\Lambda x+a))
=\exp(i \phi_{\mathcal{G}(a'+\Lambda'a,\Lambda'\Lambda)}(p,x)) \,.
\end{equation}
The equation \eqref{PoincarePhase} is fulfilled by the function
\begin{equation}
\label{PoincarePhase2}
\phi_{\mathcal{G}(a,\Lambda)}(p,x)= \kappa_G \frac{(\Lambda{p})a}{\hbar} \,,
\end{equation}
where $\kappa_G \in \RNumb$. Thus, the action of the Poincar\'e group on our
coherent states can be given by
\begin{equation}
\label{PoincareTransf2}
\hat{\mathcal{G}}(a,\Lambda) \Ket{p,x}
:= \exp \left(i\kappa_G \frac{(\Lambda{p})a}{\hbar}\right)
\Ket{\Lambda p,\Lambda x+a} \,.
\end{equation}
Two cases seem to be of particular interest: a) $\kappa_G=0$, when the phase factor does not depend on
$a,\Lambda,p,x$ and b) $\kappa_G \neq 0$, when the phase depends on
$a,\Lambda,p$. Case a) was used in \cite{IQmethod} to show covariance of the
integral quantization with respect to the Poincar\'e group. This is also true
for $\kappa_G \neq 0$. In actual applications case b) is more
useful. Using the action \eqref{PoincareTransf2} and the translation operation
from the Heisenberg--Weyl group represented by the operator
$\hat{\mathcal{U}}(0;0,a)$, one can easily show that
\begin{equation}
\label{PoincareTransf3}
\hat{\mathcal{G}}(a,\Lambda)
= \hat{\mathcal{U}}(0;0,a)\hat{\mathcal{G}}(\Lambda) \,,
\end{equation}
where the Lorentz transformation is denoted by
$\hat{\mathcal{G}}(\Lambda) \equiv
\hat{\mathcal{G}}(0,\Lambda)$. Equation \eqref{PoincareTransf3} relates the
Poincar\'e spacetime group with the Heisenberg-Weyl configuration space group.

Using the particular value of the phase factor $\kappa_G=-1/2$ one can
derive the following matrix element
\begin{eqnarray}
\label{EtakGpx}
&& \Bra{\eta_k} \hat{\mathcal{G}}(a,\Lambda) \Ket{p,x}
= \exp\left(-i\frac{(\Lambda{p})a}{2\hbar}\right) \BraKet{\eta_k}{\Lambda{p},\Lambda{x}+a}
\nonumber \\
&& =\exp\left(-i\frac{(\Lambda{p})a}{2\hbar}\right)
\exp\left(i\frac{[(\Lambda{p})-2k]a}{2\hbar}\right)
\BraKet{\eta_k}{\Lambda{p},\Lambda{x}}  \nonumber\\
&&=\exp\left(-i\frac{ka}{\hbar} \right)
\BraKet{\eta_k}{\Lambda{p},\Lambda{x}} \,,
\end{eqnarray}
where
\begin{eqnarray}
\label{EtakGpx2}
&&\BraKet{\eta_k}{\Lambda{p},\Lambda{x}} =
\int_{\RNumb^4} d^4\xi\, \eta_k(\xi)^\star
\exp\left(-i\frac{(\Lambda{p})(\Lambda{x})}{2\hbar} \right)
\exp\left(i\frac{(\Lambda{p}\xi)}{\hbar} \right) \Phi_0(\xi-\Lambda{x})
\nonumber\\
&& = \exp\left(i\frac{px}{2\hbar} \right)
\exp\left(-i\frac{k(\Lambda{x})}{\hbar} \right)
\BraKet{\eta_{k-\Lambda{p}}}{\Phi_0} \,.
\end{eqnarray}
In Eq.~\eqref{EtakGpx2} the change of variables $\xi \to \xi +\Lambda{x}$ is
performed and the Lorentz invariance of the four-vector scalar product is used.

The Poincar\'e transformation applied to the transition amplitude \eqref{Tran2}
can be written as
\begin{eqnarray}
\label{Tran3}
&& \mathcal{A}'_{m,\epsilon}
=  \Bra{p'',x''}\hat{\mathcal{G}}(a,\Lambda)^\dagger
P_{\mathcal{J}_{m,\epsilon}}
\hat{\mathcal{G}}(a,\Lambda)\Ket{p',x'} \nonumber \\
&& = \int_{\RNumb^4} d^4k
\BraKet{\eta_k}{\Lambda{p''},\Lambda{x''}}^\star
\chi(k \in \mathcal{J}_{m,\epsilon})
\BraKet{\eta_k}{\Lambda{p'},\Lambda{x'}} \,.
\end{eqnarray}
Therefore, the amplitude \eqref{Tran3} is invariant with respect to the
translation in spacetime, as the right-hand side of \eqref{Tran3} does not
depend on $a$. This important feature is independent of the form of the
fiducial vector $\Phi_0$.

The invariance of the amplitude \eqref{Tran3} with respect to the Lorentz
transformations can be achieved if the fiducial vector satisfies
$\Phi_0(\Lambda{\xi})=\Phi_0(\xi)$.  In this case, due to an invariant measure of
the scalar product in the carrier space $\StateSpace{K}$, one obtains the
following relation
\begin{equation}
\label{EtakGpx3}
\BraKet{\eta_{k-\Lambda{p}}}{\Phi_0}
= \BraKet{\eta_{\Lambda^{-1}k-p}}{\Phi_0} \,.
\end{equation}
To derive the above relation, one needs to change the integration variables
$\xi \to \Lambda{\xi}$ and use the invariance properties of the four-vector
scalar product.

Inserting the relation \eqref{EtakGpx3} into the amplitude \eqref{Tran3} and
using the Lorentz invariance of the characteristic function $\chi(k \in
\mathcal{J}_{m,\epsilon})$, one gets
\begin{eqnarray}
\label{Tran4}
&& \mathcal{A}'_{m,\epsilon}
=\Bra{p'',x''}\hat{\mathcal{G}}(a,\Lambda)^\dagger
P_{\mathcal{J}_{m,\epsilon}}
\hat{\mathcal{G}}(a,\Lambda)\Ket{p',x'} \nonumber \\
&& =\exp\left(-i\frac{p''x''-p'x'}{2\hbar} \right)
\int_{\RNumb^4} d^4k
\exp\left(i\frac{(\Lambda^{-1}k)(x''-x')}{\hbar} \right) \nonumber \\
&& \BraKet{\eta_{\Lambda^{-1}k-p''}}{\Phi_0}^\star
\chi(k \in \mathcal{J}_{m,\epsilon})
\BraKet{\eta_{\Lambda^{-1}k-p'}}{\Phi_0} \nonumber \\
&& =\Bra{p'',x''} P_{\mathcal{J}_{m,\epsilon}} \Ket{p',x'}
 = \mathcal{A}_{m,\epsilon} \,.
\end{eqnarray}
This proves that for a Lorentz-invariant fiducial vector, the transition
amplitude $\mathcal{A}_{m,\epsilon}$ is invariant with respect to the Poincar\'e
group of spacetime transformations.

We emphasise that the construction of an appropriate Lorentz-invariant fiducial vector remains an open problem requiring further
analysis.

%%%%%%%%%%%%%%%%%%%%%%%%%%%%%%%%%%%%%%%%%%%%%%%%%%%%%%%%%%%%%%%%%%%%
\section{Transition amplitude with a harmonic oscillator ground state as the
fiducial vector}

In this subsection the transition amplitude is calculated with the four
dimensional harmonic oscillator ground state \eqref{Ham17} as the fiducial
vector. This fiducial vector minimizes the Heisenberg uncertainty principle, and
it is suitable for a semiclassical description. According to previous
considerations, this amplitude is transitionally and rotationally invariant. The
rotational invariance is implied by the following fiducial vector invariance
$\Phi_0(\Lambda_R{\xi})=\Phi_0(\xi)$, where $\Lambda=\Lambda_R$
represents the spatial rotation. However, one needs to note, that this fiducial
vector breaks the Lorentz boost symmetry.

Without loss of generality, we assume that the final time components of the
momentum are such that $p_0'' \geq p_0'$.  A straightforward calculation yields
\begin{eqnarray}
\label{eq:TransAmp}
&& \mathcal{A}_{m,\epsilon} =   \exp\left( i \frac{p'\,x'-p''\,x''}{2\hbar} \right)\, \mathcal{B}
\int_{\RNumb^3} d^3_0 p\, \int_{\RNumb} dp_0\,\,  \chi(p \in
\mathcal{J}_{m,\epsilon}) \nonumber \\
&& \exp\left( \frac{i}{\hbar} p_0 (x''^{0}-x'^{0})
-\frac{1}{\hbar \lambda_0} (p_0-\bar{p}_0)^2 \right)
\exp\left(\frac{i}{\hbar} \mathbf{p} \cdot (\mathbf{x}''- \mathbf{x}')
-\frac{1}{\hbar \lambda_3} (\mathbf{p} - \bar{\mathbf p})^2 \right) ,
\end{eqnarray}
where $\bar{p}_\mu:=\frac{1}{2}(p''_\mu +p'_\mu)$. Here
$d^3_0 p :=dp_1dp_2dp_3$; the upper index denotes the dimension of the integral,
while the lower index corresponds to coordinates which are omitted from the set
$(p_0,p_1,p_2,p_3)$. The factor $\mathcal{B}$, depending only on the initial and
final states, reads
\begin{equation}
\label{eq:FTFFidVecSpc}
\mathcal{B}:= \prod_{\mu=0}^3 (\pi \hbar \lambda_\mu)^{-\frac{1}{2}}
\exp\left(- \frac{(p''_{\mu}-p'_{\mu})^2}{4\hbar \lambda_\mu}\right) \, ,
\end{equation}
where $\lambda_1=\lambda_2=\lambda_3$.

For small $\epsilon$, one can approximate $\mathcal{A}_{m,\epsilon}$ using the
mean value theorem and averaging over $p_0$. As a result one has
\begin{eqnarray}
\label{eq:AverP0TransAmp}
&& \mathcal{A}_{m,\epsilon} \approx \exp\left(i \frac{p'\,x'-p''\,x''}{2\hbar}\right)\,\,
\mathcal{B}
\int_{\RNumb^3} d^3_0 p\
\left(\sqrt{m^2+\mathbf{p}^2+\epsilon}-\sqrt{m^2+\mathbf{p}^2} \right) \nonumber \\
&&  \exp\left(\frac{i}{\hbar} [-\sqrt{m^2+\mathbf{p}^2} \, (x''^{0}-x'^{0})+
\mathbf{p} \cdot (\mathbf{x}''-\mathbf{x}')] \right) \nonumber \\
&& \exp\left( -\frac{1}{\hbar \lambda_0} [\sqrt{m^2+\mathbf{p}^2} + \bar{p}_0]^2
-\frac{1}{\hbar \lambda_3} (\mathbf{p} - \bar{\mathbf p})^2 \right) \, .
\end{eqnarray}
Performing the Taylor expansion with respect to $\epsilon$, we obtain
\begin{eqnarray}
\label{eq:AverP0TransAmp2}
&& \mathcal{A}_{m,\epsilon} \approx \epsilon \, \exp \left(i
\frac{p'\,x' - p''\,x''}{2\hbar} \right)\, \mathcal{B}
\int_{\RNumb^3} d^3_0 p\
\frac{1}{2\sqrt{m^2+\mathbf{p}^2}}  \nonumber \\
&&  \exp\left(\frac{i}{\hbar} [-\sqrt{m^2+\mathbf{p}^2} \, (x''^{0}-x'^{0})+
\mathbf{p} \cdot (\mathbf{x}''-\mathbf{x}')] \right) \nonumber \\
&& \exp\left( -\frac{1}{\hbar \lambda_0} [\sqrt{m^2+\mathbf{p}^2} + \bar{p}_0]^2
-\frac{1}{\hbar \lambda_3} (\mathbf{p} - \bar{\mathbf p})^2 \right) \, .
\end{eqnarray}
One can reduce at least one dimension in the above integral by introducing
spherical coordinates. Consider a spherical coordinate system with
$\boldsymbol\zeta = \mathbf{x}'' - \mathbf{x}'$ oriented along the $z$ axis. Let
the vector $\mathbf{p}$ have the coordinates $(|\mathbf
p|,\theta,\varphi)$.
Then
$\mathbf{p} \cdot \boldsymbol\zeta = |\mathbf p| |\boldsymbol \zeta| \cos
\theta$.
Suppose that $\bar{\mathbf p}$ is represented by the coordinates
$(|\bar{\mathbf p}|, \bar \theta, \bar \varphi)$.
Thus,
\[ \left( \mathbf{p} - \bar{\mathbf p} \right)^2 = |\mathbf p|^2 + |\bar{\mathbf
p}|^2 - 2 |\mathbf p| |\bar{\mathbf p}| \left( \cos \theta \cos \bar \theta +
\sin \theta \sin \bar\theta \cos(\varphi - \bar \varphi) \right). \]
In fact, due to axial symmetry, one could also safely set $\bar \varphi = 0$.
Equation (\ref{eq:AverP0TransAmp2}) now reads
\begin{eqnarray}
\label{eq:AverP0TransAmpSph}
&& \mathcal{A}_{m,\epsilon} \approx \epsilon \, \exp\left( i
\frac{p'\,x'-p''\,x''}{2\hbar}\right)\, \mathcal{B}
 \int_0^\infty d |\mathbf p| \int_0^{2 \pi} d \varphi \int_0^\pi d \theta
 |\mathbf p|^2 \sin \theta
\frac{1}{2\sqrt{m^2 + |\mathbf p|^2}}  \nonumber \\
&&  \exp\left(\frac{i}{\hbar} \left[ -\sqrt{m^2 + |\mathbf p|^2} \,
(x''^{0}-x'^{0})+
|\mathbf p| |\boldsymbol \zeta| \cos \theta \right] \right) \nonumber \\
&& \exp\left( -\frac{1}{\hbar \lambda_0} [\sqrt{m^2 + |\mathbf p|^2} +
\bar{p}_0]^2 \right) \nonumber \\
&& \exp\left(
-\frac{1}{\hbar \lambda_3} \left[ |\mathbf p|^2 + |\bar{\mathbf p}|^2 - 2
|\mathbf p| |\bar{\mathbf p}|
 \left( \cos \theta \cos \bar \theta + \sin \theta \sin \bar \theta \cos
 (\varphi - \bar \varphi) \right) \right] \right).
\end{eqnarray}
Now,
\begin{equation}
\int_0^{2 \pi} d \varphi \exp \left( \frac{2 |\mathbf p| |\bar{\mathbf p}| \sin
\theta \sin \bar \theta}{\hbar \lambda_3}
\cos(\varphi - \bar \varphi) \right) = 2 \pi I_0 \left( \frac{2 |\mathbf p|
|\bar{\mathbf p}| \sin \theta \sin \bar
 \theta}{\hbar \lambda_3} \right),
 \end{equation}
where $I_0$ denotes the modified Bessel function of the first kind. This gives
\begin{eqnarray}
&& \mathcal{A}_{m,\epsilon} \approx \epsilon \, \pi \exp \left( i
\frac{p'\,x'-p''\,x''}{2\hbar} \right) \mathcal{B}
\int_0^\infty d |\mathbf p| \int_0^\pi d \theta |\mathbf p|^2 \sin \theta
\frac{1}{\sqrt{m^2 + |\mathbf p|^2}} \nonumber \\
&&  \exp\left(\frac{i}{\hbar} \left[ -\sqrt{m^2 + |\mathbf p|^2} \,
(x''^{0}-x'^{0})+
|\mathbf p| |\boldsymbol \zeta| \cos \theta \right] \right) \nonumber \\
&& \exp\left( -\frac{1}{\hbar \lambda_0} [\sqrt{m^2 + |\mathbf p|^2} +
\bar{p}_0]^2 \right) \nonumber \\
&& \exp\left(
-\frac{1}{\hbar \lambda_3} \left[ |\mathbf p|^2 + |\bar{\mathbf p}|^2 - 2
|\mathbf p| |\bar{\mathbf p}|
\left( \cos \theta \cos \bar \theta \right) \right] \right) I_0 \left( \frac{2
|\mathbf p| |\bar{\mathbf p}|
\sin \theta \sin \bar \theta}{\hbar \lambda_3} \right).
\label{eq:AverP0TransAmpSphBessel}
\end{eqnarray}
If $\boldsymbol\zeta$ and $\bar{\mathbf p}$ are oriented in the same direction,
we have $\bar \theta = 0$ and
\begin{equation}
I_0\left( \frac{2 |\mathbf p| |\bar{\mathbf p}| \sin \theta \sin \bar
\theta}{\hbar \lambda_3} \right) = 1.
\end{equation}
In this case the integral over $\theta$ can be evaluated analytically. We get
\begin{eqnarray}
\mathcal{A}_{m,\epsilon} & \approx & \epsilon \, \pi \exp \left( i
\frac{p'\,x'-p''\,x''}{2\hbar} \right) \, \mathcal{B}
 \int_0^\infty d |\mathbf p| \int_0^\pi d \theta |\mathbf p|^2 \sin \theta
\frac{1}{\sqrt{m^2 + |\mathbf p|^2}} \nonumber \\
&&  \exp\left(\frac{i}{\hbar} \left[ -\sqrt{m^2 + |\mathbf p|^2} \,
(x''^{0}-x'^{0})+
|\mathbf p| |\boldsymbol \zeta| \cos \theta \right] \right) \nonumber \\
&& \exp\left( -\frac{1}{\hbar \lambda_0} [\sqrt{m^2 + |\mathbf p|^2} +
\bar{p}_0]^2 \right) \nonumber \\
&& \exp\left(
-\frac{1}{\hbar \lambda_3} \left[ |\mathbf p|^2 + |\bar{\mathbf p}|^2 - 2
|\mathbf p| |\bar{\mathbf p}| \cos \theta \right] \right) \nonumber \\
& = & \epsilon \, \pi \exp \left( i \frac{p'\,x'-p''\,x''}{2\hbar} \right) \,
\mathcal{B}
 \int_0^\infty d |\mathbf p| \frac{|\mathbf p|^2}{\sqrt{m^2 + |\mathbf p|^2}}
 \nonumber \\
&& \exp\left( -\frac{1}{\hbar \lambda_0} [\sqrt{m^2 +
 |\mathbf p|^2} + \bar{p}_0]^2 \right) \nonumber \\
&&  \exp\left(\frac{i}{\hbar} \left[ -\sqrt{m^2 + |\mathbf p|^2} \,
(x''^{0}-x'^{0}) \right] \right) \exp\left(
-\frac{1}{\hbar \lambda_3} \left[ |\mathbf p|^2 + |\bar{\mathbf p}|^2 \right]
\right)  \nonumber \\
&&  \frac{2}{\left( \frac{i}{\hbar} |\mathbf p| |\boldsymbol \zeta| +
\frac{2}{\hbar \lambda_3} |\mathbf p| |\bar{\mathbf p}|
 \right)} \sinh \left( \frac{i}{\hbar} |\mathbf p| |\boldsymbol \zeta| +
 \frac{2}{\hbar \lambda_3} |\mathbf p| |\bar{\mathbf p}| \right).
\label{eq:AverP0TransAmpSphColinar}
\end{eqnarray}

%%%%%%%%%%%%%%%%%%%%%%%%%%%%%%%%%%%%%%%%%%%%
\section{Quantum evolution of a test particle}
%%%%%%%%%%%%%%%%%%%%%%%%%%%%%%%%%%%%%%%%%%%%%

In this section we investigate basic phenomena related to the motion of a test
particle, predicted by our model.  We start by analyzing the free motion
determined by the probability amplitude \eqref{Tran2}. As a second application,
we consider a quantum random walk in the Minkowski spacetime. We conclude with
an analysis of interference patterns occurring in the standard double slit
experiment.

%The probability of quantum evolution of a test particle is calculated by using
%the formula
%
%\begin{equation}\label{App1}
%\mathcal{A}_m := \lim_{\epsilon\rightarrow 0}
%\frac{\mathcal{A}_{m,\epsilon}}{\epsilon}
%\end{equation}
%
%for a particle, and
%
%\begin{equation}\label{App2}
% \mathcal{A}_0 :=  \lim_{m \rightarrow 0} \mathcal{A}_m
%\end{equation}
%
%for a photon.

%%%%%%%%%%%%%%%%%%%%%%%%%%%%%%%%%%%%%%%%%%%%%%%
%\subsection{Quantum evolution of test particle}
\subsection{Free motion of a test particle}
\label{sec:Quantum_evolution}
%%%%%%%%%%%%%%%%%%%%%%%%%%%%%%%%%%%%%%%%%%%%%%%

In our model, a free motion of a test particle in the Minkowski spacetime is
determined by the transition amplitude \eqref{Tran2} and the resulting
transition probability.

\begin{figure}
\begin{center}
\includegraphics[width=0.49\textwidth]{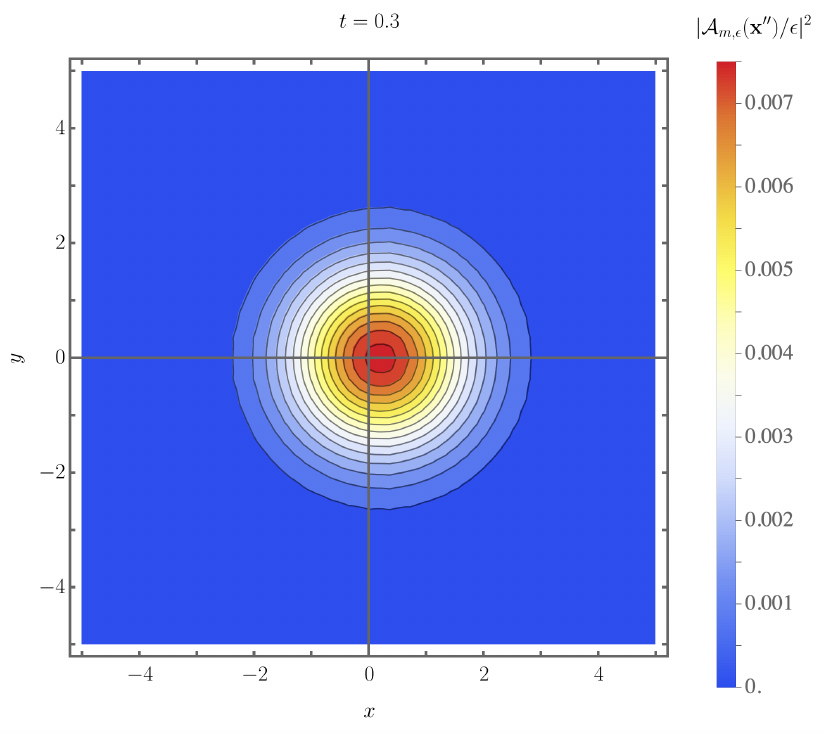}\includegraphics[width=0.49\textwidth]{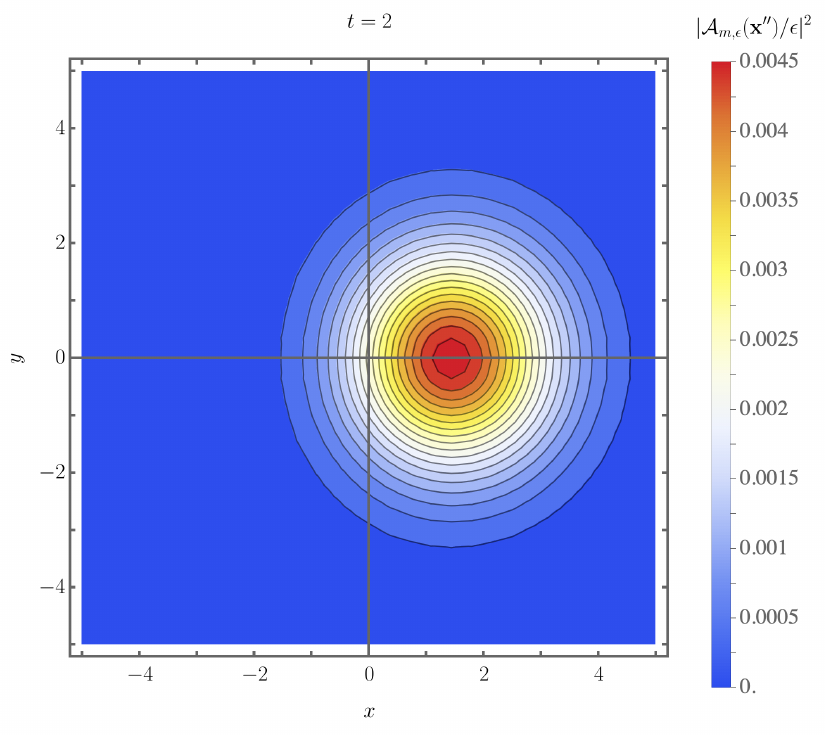}
\end{center}
\caption{
\label{Fig:P-stwa}
The density
$|\mathcal{A}_{m,\epsilon}(\mathbf{x}'')/\epsilon|^2$ computed according to
(\ref{eq:AverP0TransAmpSphColinar}) as a function of $\mathbf{x}''$ for $t=0.3$ (left) and $t =2$ (right).
Here $m=1$, $\mathbf{p}_0=(1,0,0)$,
$\mathbf{x}'=(0,0,0)$, $p_0' =-\sqrt{m^2+ (\mathbf{p}')^2}$,
$\lambda_0 =3\lambda_3 =3$. Both plots show the graphs in the $xy$-plane of the vector $\mathbf x''$.}
\end{figure}

\begin{figure}
\begin{center}
\includegraphics[width=0.49\textwidth]{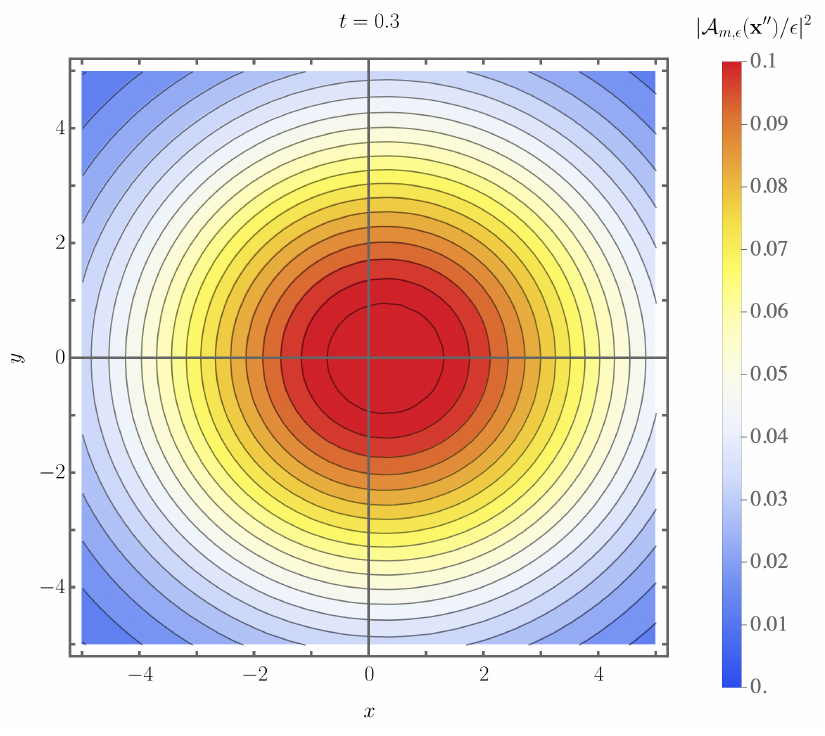}\includegraphics[width=0.49\textwidth]{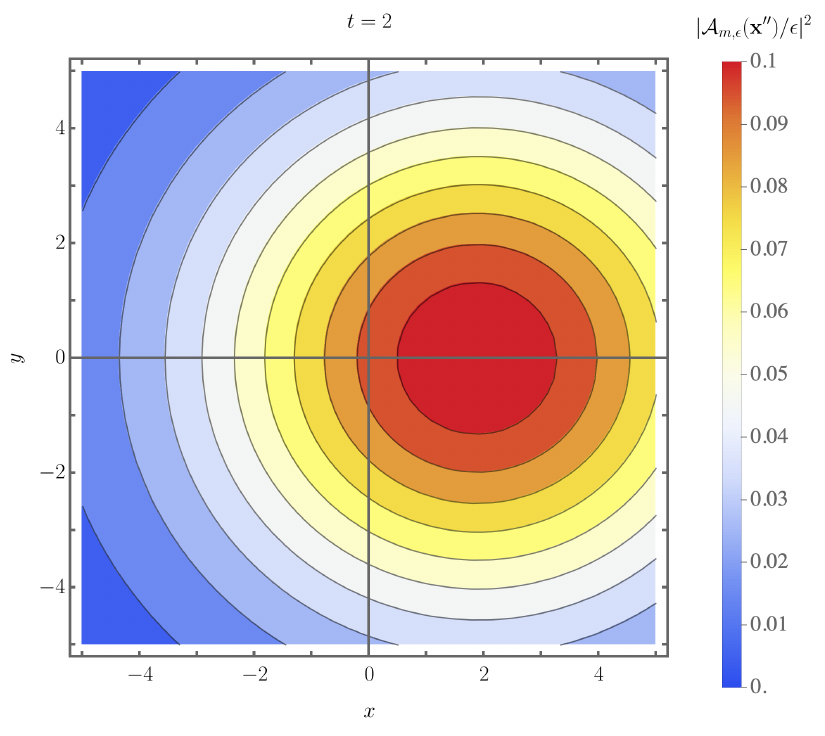}
\end{center}
\caption{
\label{Fig:P-stwa01}
Same as in Fig.\ \ref{Fig:P-stwa}, but for $\lambda_0 =3\lambda_3 =0.3$.}
\end{figure}

\begin{figure}
\begin{center}
\includegraphics[width=0.49\textwidth]{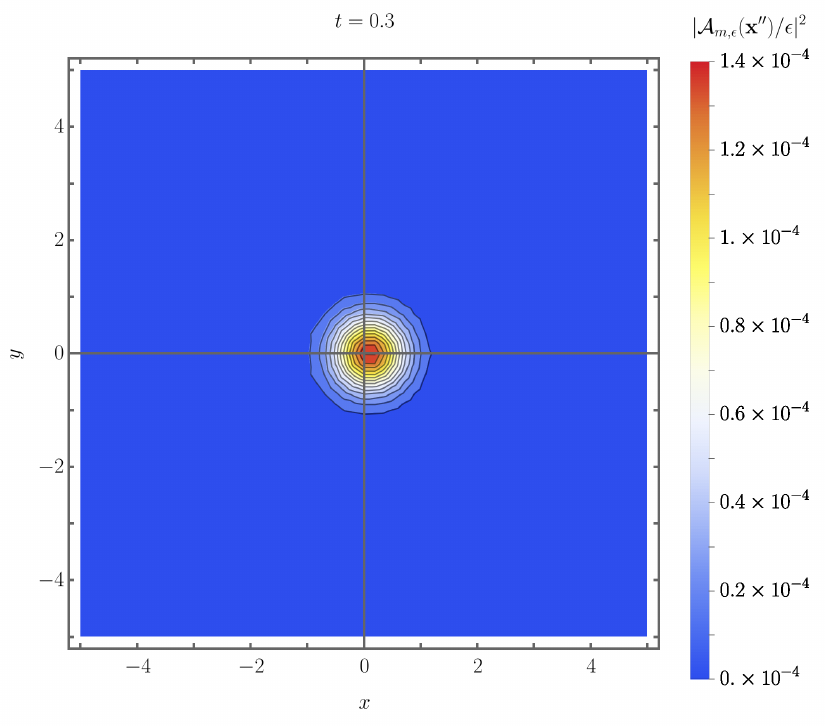}\includegraphics[width=0.49\textwidth]{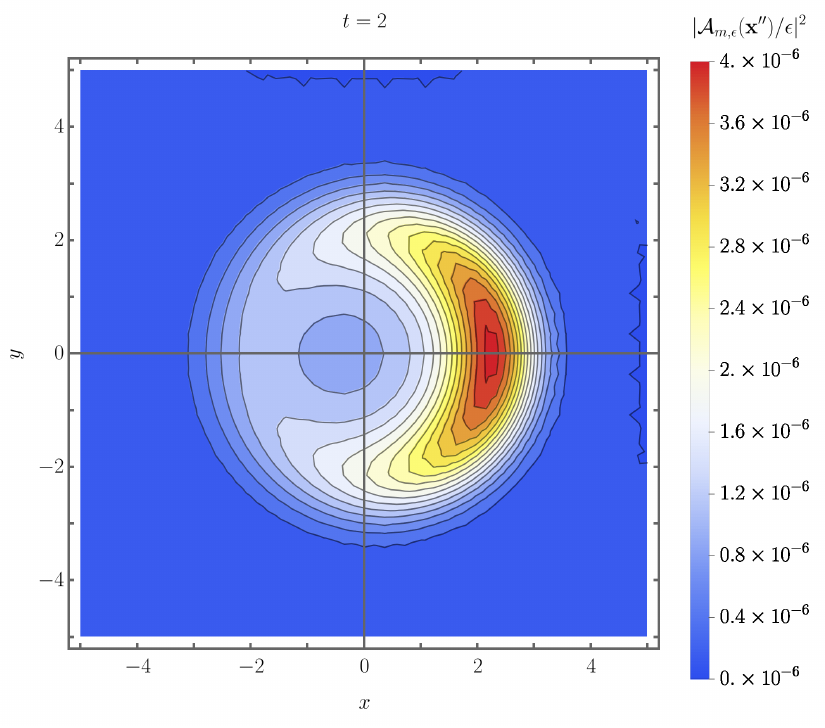}
\end{center}
\caption{
\label{Fig:P-stwa10}
Same as in Fig.\ \ref{Fig:P-stwa}, but for $\lambda_0 =3\lambda_3 =30$.}
\end{figure}

Figures \ref{Fig:P-stwa}, \ref{Fig:P-stwa01}, and \ref{Fig:P-stwa10} present the
density plots of $|\mathcal{A}_{m,\epsilon}/\epsilon|^2$ understood as a
function of the final positions $\mathbf{x}''$ for two different times
$t:={x\,}''^0$, and computed according to Eq.\
(\ref{eq:AverP0TransAmpSphColinar}).  Figures \ref{Fig:P-stwa},
\ref{Fig:P-stwa01}, and \ref{Fig:P-stwa10} were obtained for
$\lambda_0 =3\lambda_3 =3$, $\lambda_0 =3\lambda_3 =0.3$, and
$\lambda_0 =3\lambda_3 =30$, respectively.  In all figures, $x’=(0,0,0,0)$ and
$p’={p\,}''$ are fixed and placed on the mass shell, where $m=1$,
$\mathbf{p}'=(1,0,0)$ and $p’_0 = -\sqrt{m^2+\mathbf{p}'^2}$. The plots present
the probability distribution of a transition of a particle from the space point
$(0,0,0)$ to $\mathbf{x}''$ in time $t$. The momenta $p’$ and ${p\,}''$ are
defined by Eq.\ (\ref{HamSol}).  For all $t$ we obtain an axially symmetric
distribution, where the symmetry axis is aligned with the direction of the
spatial momentum (the plots in the $xz$-plane would look the same). Figures
\ref{Fig:P-stwa}, \ref{Fig:P-stwa01}, and \ref{Fig:P-stwa10} differ in the width
of the distributions.  For small $\lambda_\mu$ (Fig. \ref{Fig:P-stwa01}), the
position localization of the particle has a large uncertainty, which implies
that the momentum components are well defined. This fact can be observed in the
unchanged width of the probability distribution for greater time. On the other
hand, for a large $\lambda_\mu$ (Fig. \ref{Fig:P-stwa10}) the position
distribution exhibits a sharp maximum. Thus, the momenta are characterized by
large uncertainties, leading to a fast increase of the smearing.  In all
pictures, the probability distribution follows the classical geodesic path---its
maximum remain located at the geodesic trajectory. For larger $t$ the
probability distribution gets wider, and it expands in the transverse plane.
This means that transitions corresponding to larger times become more deflected
from the classical trajectory, i.e., they are characterized by a larger
spread. On the other hand, the maximum of the probability distribution becomes
smaller with growing time.

%%%%%%%%%%%%%%%%%%%%%%%%
%%%%%%%%%%%%%%%%%%%%%%%
%%%%%%%%%%%%%%%%%%%%%%%%%%
%%%%%%%%%%%%%%%%%%%%%%%%%

There exists a visible correspondence between the quantum nature of our
description and the related classical one.  This correspondence is implied by
the structure of the mass shell \eqref{eq:ProjMassLayer}, constructed from the eigenstates of the geodesic
Hamiltonian. The transition operator
$P_{\mathcal{J}_{m,\epsilon}}$ distinguishes appropriate spatial positions of
the particle as the ones with greater probability, and it introduces the time
dependence in the probability distribution.

\begin{figure}
\begin{center}
\includegraphics[width=0.49\textwidth]{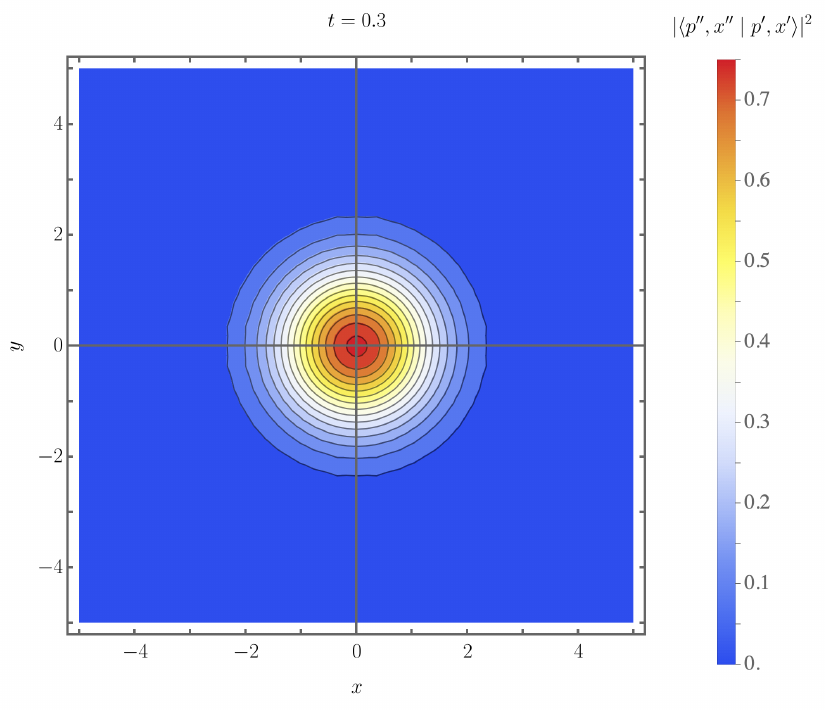}\includegraphics[width=0.49\textwidth]{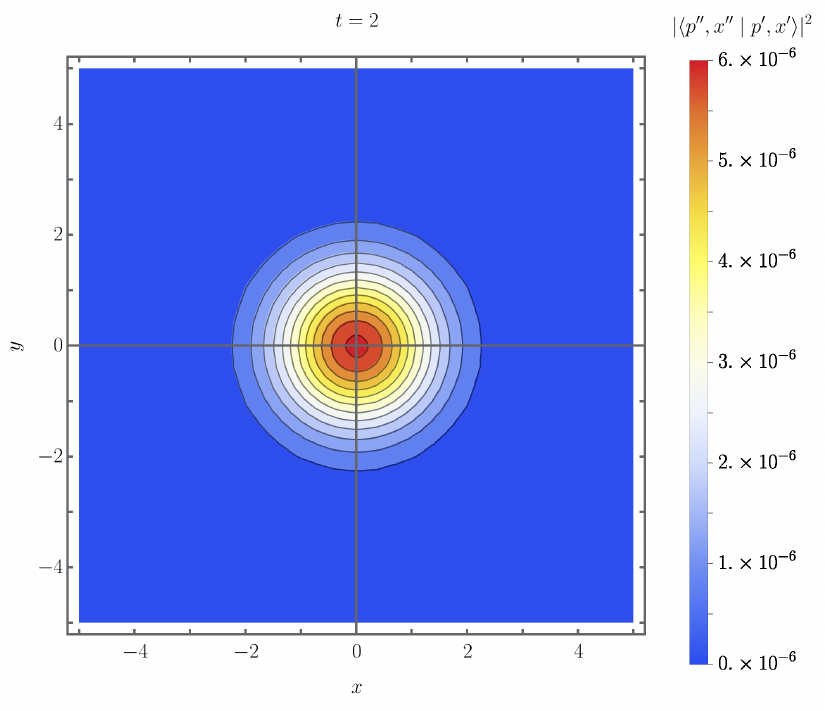}
\end{center}
\caption{
\label{Fig:PrProb}
The density $|\BraKet{p'',x''}{p',x'}|^2$ computed according to Eq.\ (\ref{cohbraket})
as a function of $\mathbf{x}''$ for $t=0.3$ (left) and $t=2$ (right). Here $m=1$, $\mathbf{p}_0=(1,0,0)$, $\mathbf{x}'=(0,0,0)$,
$p_0' =-\sqrt{m^2+ (\mathbf{p}')^2}$,
$\lambda_0 =3\lambda_3 =3$. Both plots show the graphs in the $xy$-plane of the vector $\mathbf x''$.}
\end{figure}
%%%%%%%%%%%%

Figure \ref{Fig:PrProb} presents the overlap $|\BraKet{p'',x''}{p',x'} |^2$,
computed in Appendix \ref{Prob}. It is proportional to the spherically symmetric
Gaussian distribution \eqref{cohbraket}, but it does not correspond to an
evolution along a geodesic trajectory. The obtained probability densities
decrease rapidly with time $t$ and distance $|\mathbf{x}''|$, much faster than
the densities shown in Fig.\ \ref{Fig:P-stwa}. Figure \ref{Fig:PrProb}
illustrates the existence of correlations among all points of the quantum
configuration space, which is a specific feature of the integral
quantization. In the canonical quantization, the ``corresponding'' overlap would
vanish, due to the orthogonality of states. This feature represents one of main
differences between integral and canonical quantizations.

%%%%%%%%%%%%%%%%%%%%%%%%%%%%%%%%%
\subsection{Quantum random walks}
%%%%%%%%%%%%%%%%%%%%%%%%%%%%%%%%%
%%%
\begin{figure}
\begin{center}
\includegraphics[width=0.5\textwidth]{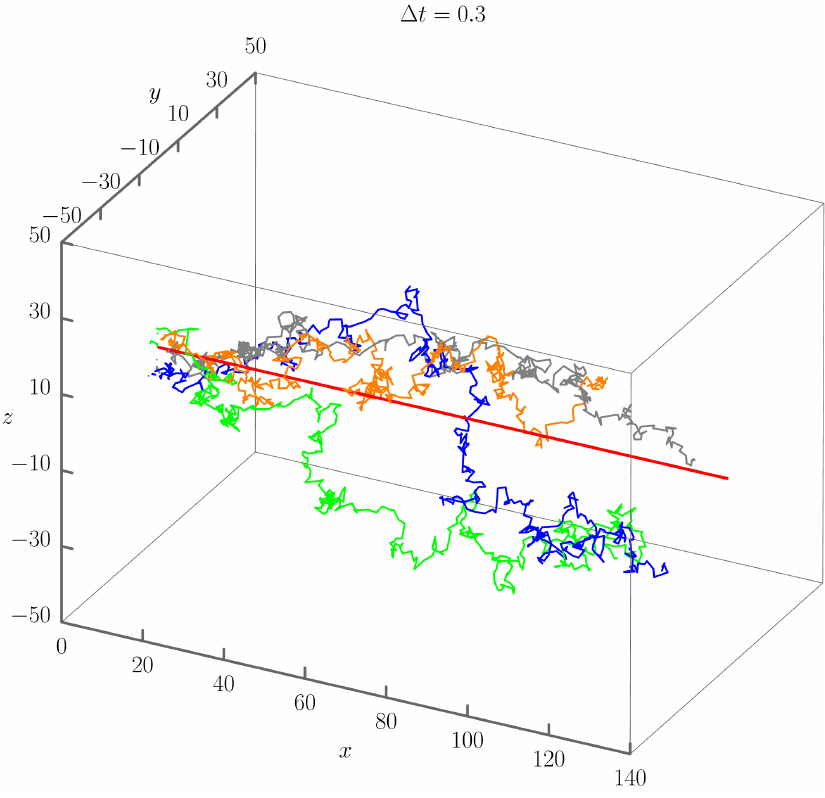}
\includegraphics[width=0.5\textwidth]{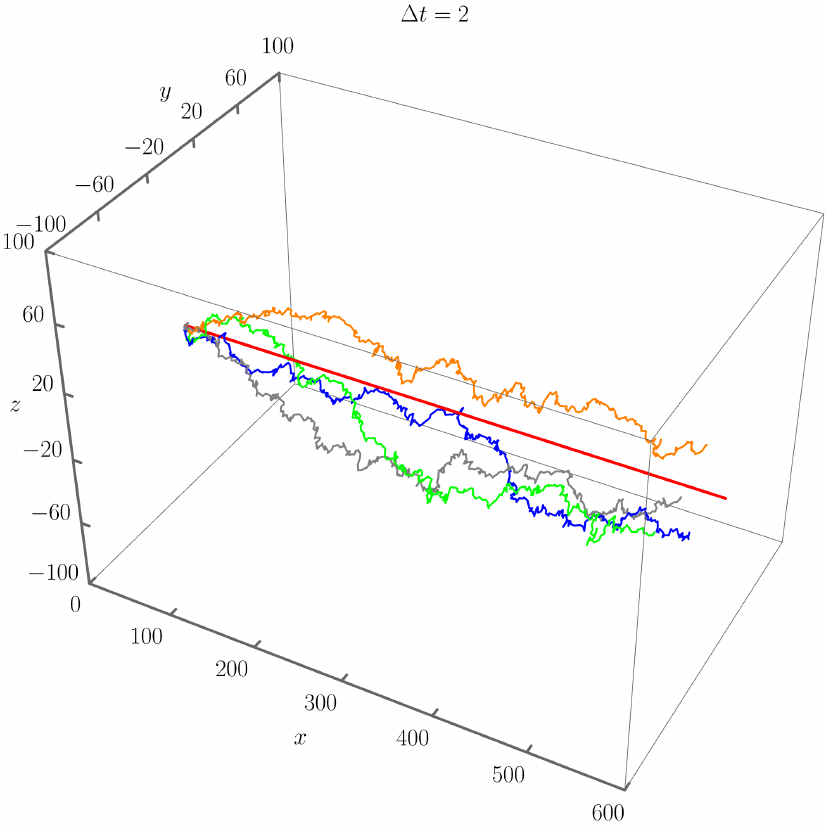}
\end{center}
\caption{
\label{Fig:RWalk}
Sample stochastic random walks of a particle, where $\Delta t=0.3$ and $\Delta t=2$ are
time periods between ``measurements'' of the particle positions. The probability
distributions are the same as presented in Fig.~\ref{Fig:P-stwa}. Thus, $m=1$,
$\mathbf{p}'=(1,0,0)$, $\mathbf{x}'=(0,0,0)$, ${p'}_0=-\sqrt{m^2+\mathbf{p}'^2}$,
$\lambda_0=3\lambda_3=3$. Each stochastic trajectory consists of 500 points.}
\end{figure}
%%%

As another visualization of the properties of the obtained probability
distributions, we compute examples of a quantum random walk of a particle.

Figure \ref{Fig:RWalk} represents stochastic trajectories of particles by
polylines with vertices coinciding with positions of a given particle in
every subsequent step. The trajectories are constructed with help of a computer
generator of random numbers implemented in Mathematica \cite{M10} and combined
with the probability distribution resulting from \eqref{eq:TransAmp}.
Physically, one can interpret such trajectories as a result of some sort of
``experiment'', consisting in finding the position of the quantum particle after
each time interval $\Delta t$.  This experiment, repeated in  subsequent time
intervals $\Delta t$, corresponds to the transition
$|p', x'\rangle \rightarrow |p'', x'' \rangle$ with $p'=p''$, i.e., satisfying
the condition that the expectation values of the initial and final four-momenta
are the same. The vertices of a given polyline represent the spatial positions of
the particle and the line segments of the polyline are used to ascribe the time
sequence to subsequent experiments.

Figure \ref{Fig:RWalk} presents sample trajectories for $\Delta t=0.3$ and
$\Delta t = 2$ (initial conditions for probability distributions are the same as
given in Fig.\ \ref{Fig:P-stwa}). The red line represents the classical
geodesic. In each case, one can observe that on average the particle position
remains close to the classical path. For larger $\Delta t$ the quantum particle
has longer ``quantum'' jumps, implying a tendency to exhibit larger deviation
from the classical geodesic. This is a consequence of the position of a maximum
and the variance of the suitable probability distribution.

%%%%%%%%%%%%%%%%%%%%%%%%%%%%%%%%%%
\subsection{Interference patterns}
%%%%%%%%%%%%%%%%%%%%%%%%%%%%%%%%%%

\begin{figure}
\begin{center}
\includegraphics[width=0.8\textwidth]{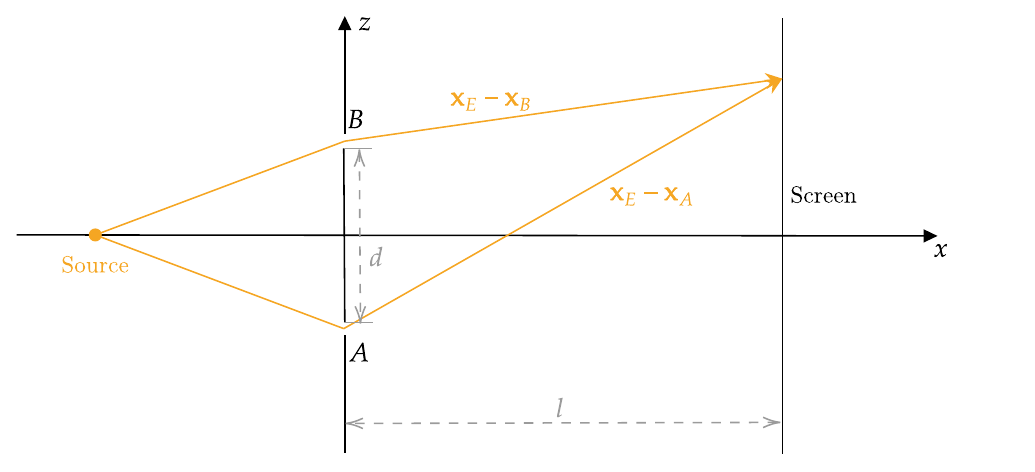}
\end{center}
\caption{\label{fig:interferencesetup} A simplified diagram of the double-slit
experiment setup. Orange lines represent classical trajectories passing through
slits $A$ and $B$, separated by a distance $d$ and located at a distance $l$
from the screen.}
\end{figure}

As a last application of our model, we will discuss a standard double-slit
interference predicted by our quantization scheme. Consider an experimental setup
illustrated in Figure \ref{fig:interferencesetup}. A source and a screen are
separated by a plate with two slits. The distance between the slits and the
distance between the plate and the screen are denoted by $d$ and $l$,
respectively. We assume that $d \ll l$.

In classical physics, the slits are often modeled as focal points (sources) of
spherical or cylindrical waves. In quantum mechanics, the slits should be treated as physical
objects interacting with particles passing through them and changing their quantum states.

As a first and natural approach, we used coherent states to represent quantum states
of a particle at the source, the slits, and the screen. For a reasonable distance between the slits,
the coherent states generated from the harmonic oscillator vacuum fiducial vector are approximately
orthogonal. Assuming this approximation, one can use the following standard combination of transition
amplitudes \eqref{eq:TransAmp} to calculate the probability of detecting a particle on the screen $E$:
\begin{eqnarray}
\label{InterfProbCS}
&& M=|\Bra{p_E,x_E} P_{\mathcal{J}_{m,\epsilon}} \Ket{p_{A},x_{A}}
\BraKet{p_{A},x_{A}}{p_{A'},x_{A'}}
\Bra{p_{A'},x_{A'}}  P_{\mathcal{J}_{m,\epsilon}} \Ket{p_{Z},x_{Z}}
\nonumber\\
&& +
\Bra{p_E,x_E} P_{\mathcal{J}_{m,\epsilon}} \Ket{p_{B},x_{B}}
\BraKet{p_{B},x_{B}}{p_{B'},x_{B'}}
\Bra{p_{B'},x_{B'}}  P_{\mathcal{J}_{m,\epsilon}} \Ket{p_{Z},x_{Z}}|^2 \,.
\end{eqnarray}
Here $Z$ denotes the source of particles, $A',B'$ and $A,B$ denote ingoing
and outgoing states of particles passing through the slits, and $E$ denotes
the screen. In this formalism, the amplitude
$\BraKet{p_{S},x_{S}}{p_{S'},x_{S'}}$, where $S=A,B$, describes the behaviour
of a particle inside the slit, and it can be regarded as corresponding to a random transition
between
coherent states due to their overlaps. We should note that the choice of this
amplitude can be understood as representing a physical model of the slit, and
different physical realizations of the slits would, in general, lead to different
amplitudes.

\begin{figure}
\begin{center}
\includegraphics[width=0.6\textwidth]{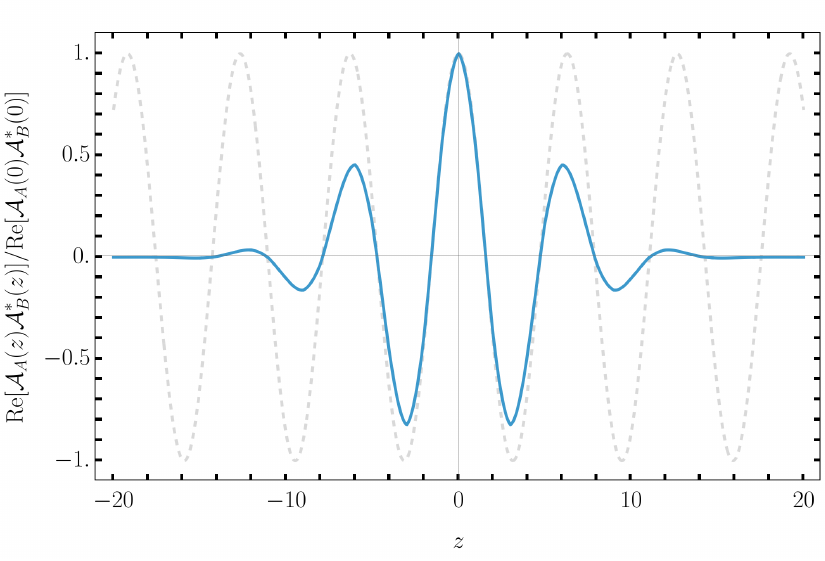}
\end{center}
\caption{\label{fig:InterferenceTermKoher}
Interference term $\mbox{Re}(\mathcal{A}_A(z)\mathcal{A}^\star_B(z))/\mbox{Re}(\mathcal{A}_A(0)\mathcal{A}^\star_B(0))$,
where $\mbox{Re}(\mathcal{A}_A(0)\mathcal{A}^\star_B(0))\simeq 3.42 \times 10^{-13}$,
as a function of the position on the screen $z$ (blue curve).
Here $\hbar=1$, $m=1$, $3\lambda_3=\lambda_0=3$,  $l=100$, $|\mathbf{p}|=20$, $d=5$;
the distance between the source and the slits is equal to 20; the time of travel between the source
and the slits is 20, the time of travel between the slits and the screen equals 100.
The gray dotted curve represents plot of $\cos(|\mathbf{p}|(|\Delta\mathbf{x}_{EA}|-|\Delta\mathbf{x}_{EB}|)/\hbar)$.
}
\end{figure}

Unfortunately, the integrals in the expression \eqref{InterfProbCS} cannot be
expressed by elementary functions. In the asymptotic region, i.e., when the distance
between the slits and the screen is much larger than the distance between the slits,
the numerical results reproduce well-known relations among interference fringes.
Figure \ref{fig:InterferenceTermKoher} presents a numerical calculation of
the interference term $\mbox{Re}[\mathcal{A}_A(z)\mathcal{A}^\star_B(z)]$
deriving from  (\ref{InterfProbCS}) (see also Eq.\ (\ref{eq:InterM}) below) as a function of the position on a screen $z$, where
\begin{equation}
\mathcal{A}_S(z)=
\Bra{p_E,x_E}P_{\mathcal{J},m}\Ket{p_{S’},x_{S’}}
\BraKet{p_{S’},x_{S’}}{p_{S},x_{S}}
\Bra{p_S,x_S}P_{\mathcal{J},m}\Ket{p_Z,x_Z}\, .
\end{equation}
The geometric layout is delineated in Fig.\ \ref{fig:interferencesetup},
where  $x_{A'} = x_A$ and $x_{B'} = x_B$. The directions of the spatial parts of
the momenta correspond to appropriate geometrical directions, i.e., $\mathbf{p}_{A'}$
is proportional to $\mathbf{x}_A-\mathbf{x}_Z$ and $\mathbf{p}_A$ is proportional
to $\mathbf{x}_E-\mathbf{x}_A$ (similarly for $\mathbf{p}_{B'}$ and $\mathbf{p}_B$).
The norms of the spatial parts of the momenta are all equal to $|\mathbf{p}|$, and the time components
of all of the momenta are equal
to $-\sqrt{m^2+|\mathbf{p}|^2}$. The periodicity of the result remains in agreement with the classical expectation:
the locations of the interference maxima coincide with the maxima of
$\cos(|\mathbf{p}|(|\Delta\mathbf{x}_{EA}|-|\Delta\mathbf{x}_{EB}|)/\hbar)$,
where $\Delta\mathbf{x}_{ES}=\mathbf{x}_E-\mathbf{x}_S$.

To provide a deeper analytic insight in the non-asymptotic regime, we consider a simpler model
in which the slits are represented by the eigenstates of the position operator. This representation
mimic to some extent the classical approach with slits treated as sources of spherical waves in which
no particular direction of the spatial momentum is distinguished.

As before, we compute transition amplitudes between
position states associated with the slits and coherent states characterized by a
given momentum and a location corresponding to the observation point at the
screen. The probability of detecting a particle on the screen is obtained by
adding the amplitudes corresponding to both slits and computing the square of
the absolute value of the result. Interestingly, this model allows us to
calculate the ``half-way'' amplitude (from the slits to the screen only)
to obtain a correct distribution of the interference fringes
\begin{equation}
\label{InterfProbX}
M=|\Bra{p_E,x_E} P_{\mathcal{J}_{m,\epsilon}} \Ket{x_{A}}
+\Bra{p_E,x_E} P_{\mathcal{J}_{m,\epsilon}} \Ket{x_{B}}|^2\,.
\end{equation}
It was shown in \cite{IQmethod} that the eigenstates $\Ket{x}$ of the position
operator $\hat{x}^\mu$ satisfy
\begin{eqnarray}
\BraKet{x'}{p,x} & = &\exp\left(-i\frac{p_\mu x^\mu}{2\hbar}\right)
\exp\left(i\frac{p_\mu x'^\mu}{\hbar}\right)\Phi_0(x'-x)\, ,\\
\BraKet{x'}{\eta_p} & = &\frac{1}{(2\pi\hbar)^2}\exp\left(i\frac{p_\mu x'^\mu}{\hbar}\right)\, .
\end{eqnarray}
The amplitude of the transition from state $\Ket{x'}$ to $\Ket{p'',x''}$ can be
computed, analogously to Eq.\ (\ref{Tran2}), as
\begin{eqnarray}
\lefteqn{\mathcal{A}_{x'\rightarrow (p'',x'')}=\Bra{p'',x''}P_{\mathcal{J}_{m,\epsilon}}\Ket{x'}
=\int_{\RNumb^4}d^4p\frac{1}{\pi\hbar\sqrt[4]{\lambda_0\lambda_3^3}}\exp\left(-i\frac{(p''-2p)x''}{2\hbar}\right)} \nonumber\\
&&\exp\left(-\frac{(p_0-p''_0)^2}{2\hbar\lambda_0}\right)
\exp\left(-\frac{(\mathbf{p}-\mathbf{p\;}'')^2}{2\hbar\lambda_3}\right)\chi(p\in\mathcal{J}_{m,\epsilon})
\frac{1}{(2\pi\hbar)^2}\exp\left(-i\frac{px'}{\hbar}\right) \nonumber\\
& = &
\exp\left(i\frac{p''x''}{2\hbar}\right)\exp\left(-i\frac{p'' x'}{\hbar}\right)
\frac{1}{4(\pi\hbar)^3\,\sqrt[4]{\lambda_0\lambda_3^3}}\nonumber\\
&&
\int_{\RNumb^4}d^4 p\;\exp\left(i\frac{p(x''-x')}{\hbar}\right)\exp\left(-\frac{p_0^2}{2\hbar\lambda_0}\right)
\exp\left(-\frac{\mathbf{p\,}^2}{2\hbar\lambda_3}\right)\chi(p+p''\in\mathcal{J}_{m,\epsilon}) \nonumber\\
& = &\exp\left(i\frac{p''x''}{2\hbar}\right)\exp\left(-i\frac{p'' x'}{\hbar}\right) X_{x'\rightarrow (p'',x'')} \, ,
\end{eqnarray}
where
\begin{eqnarray}
&&X_{x'\rightarrow (p'',x'')}=\frac{1}{4(\pi\hbar)^3\,\sqrt[4]{\lambda_0\lambda_3^3}}\nonumber\\
&&
\int_{\RNumb^4}d^4 p\;\exp\left(i\frac{p(x''-x')}{\hbar}\right)\exp\left(-\frac{p_0^2}{2\hbar\lambda_0}\right)
\exp\left(-\frac{\mathbf{p\,}^2}{2\hbar\lambda_3}\right)\chi(p+p''\in\mathcal{J}_{m,\epsilon})\, .
\end{eqnarray}
The interference pattern has a form
\begin{eqnarray}
M & = &\left|\mathcal{A}_{x_A\rightarrow (p_E,x_E)}+\mathcal{A}_{x_B\rightarrow (p_E,x_E)}\right|^2 \nonumber\\
& = &|X_{x_A\rightarrow (p_E,x_E)}|^2+|X_{x_B\rightarrow (p_E,x_E)}|^2 \nonumber\\
&& +2\mbox{Re}\left[\exp\left(i\frac{p_E(x_B- x_A)}{\hbar}\right)X_{x_A\rightarrow (p_E,x_E)}X^\star_{x_B\rightarrow (p_E,x_E)}\right]\, .
\label{eq:InterM}
\end{eqnarray}

\begin{figure}
\begin{center}
\includegraphics[width=0.8\textwidth]{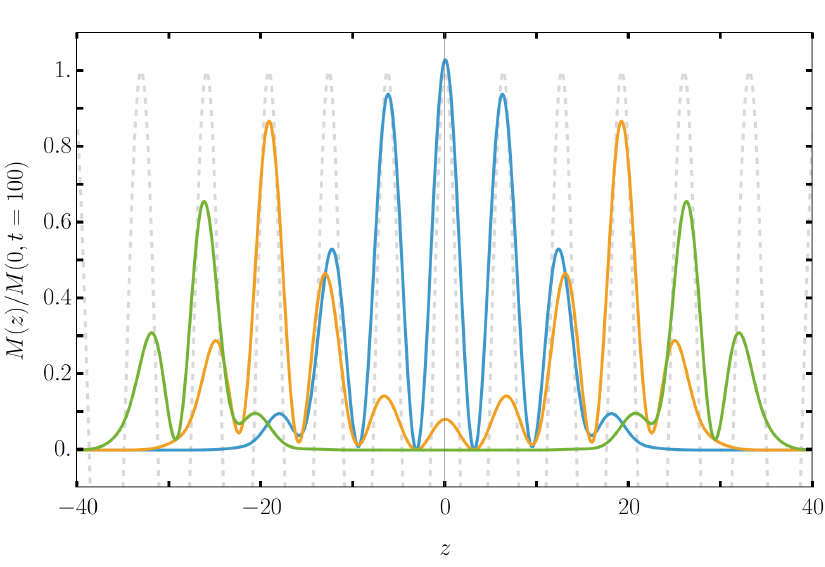}
\end{center}
\caption{
\label{Fig:Int}
Interference pattern $M(z)/M(z=0,t=100)$, where $M(z=0,t=100)\simeq 5.3 \times 10^{-7}$, as a function of the position $z$  on the screen.
Here $\hbar=1$, $m=1$, $3\lambda_3=\lambda_0=3$,  $l=100$, $|\mathbf{p}_E|=20$, $d=5$.
Different colors represent different times of the measurement: blue, orange, and green correspond to $t=100$, $t=102$, and $t=104$, respectively.
The gray dotted curve shows a graph of $\cos(|\mathbf p_E| z d/( \hbar \sqrt{l^2 + z^2}))$.
}
\end{figure}

Let us choose the origin of the coordinate system at the midpoint between the
slits (see Fig. \ref{fig:interferencesetup}).  The positions of the slits $x_A$,
$x_B$, the momentum at the screen $p_E$, and the location of the observation
point can be expressed as
\begin{eqnarray}
&& x_A=\left(0,0,0,-\frac{d}{2}\right)\, ,\quad x_B=\left(0,0,0,\frac{d}{2}\right),
\label{eq:IntVec1}\\
&&p_E=\left(-\sqrt{m^2+|\mathbf{p}_E|^2},\frac{|\mathbf{p}_E|l}{\sqrt{l^2+z^2}},0,\frac{|\mathbf{p}_E|z}{\sqrt{l^2+z^2}}\right)\, , \quad x_E=(t,l,0,z).
\label{eq:IntVec2}
\end{eqnarray}
It follows that
\begin{equation}
p_E(x_B-x_A)=\frac{|\mathbf{p}_E|zd}{\sqrt{l^2+z^2}} \, .
\end{equation}
This result can be compared with a classical calculation in which the amplitudes
$\mathcal A_{x_A \to (p_E,x_E)}$ and $\mathcal A_{x_B \to (p_E,x_E)}$ are
replaced by $A \exp(i p_A ( x_E - x_A)/\hbar)$ and
$B \exp(i p_B ( x_E - x_B)/\hbar)$, where $A$ and $B$ are functions of
$|\mathbf x_E - \mathbf x_A|$ and $|\mathbf x_E - \mathbf x_B|$,
respectively. Here the momenta $p_A$ and $p_B$ are given by
\begin{equation}
    p_A = (p_0,\mathbf p_A), \quad p_B = (p_0, \mathbf p_B),
\end{equation}
and we assume that $|\mathbf p_A| = |\mathbf p_B| =: |\mathbf p|$. The vectors
$\mathbf p_A$ and $\mathbf p_B$ are proportional to
$\mathbf x_E - \mathbf x_A$ and $\mathbf x_E - \mathbf x_B$, respectively. This
yields
\begin{eqnarray}
    M & = & \left|A \exp\left(i \frac{p_A ( x_E - x_A)}{\hbar}\right) +  B \exp\left(i \frac{p_B ( x_E - x_B)}{\hbar}\right)\right|^2 \nonumber \\
    & = & |A|^2 + |B|^2 + 2 \mathrm{Re} \left[ A B^\star \exp \left( i \frac{|\mathbf p| (|\mathbf x_E - \mathbf x_A| - |\mathbf x_E - \mathbf x_B|)}{\hbar} \right) \right] \nonumber \\
    & = &|A|^2 + |B|^2 + 2 |A| |B| \cos \left(\frac{ |\mathbf p| (|\mathbf x_E - \mathbf x_A| - |\mathbf x_E - \mathbf x_B|)}{\hbar} + \theta_A - \theta_B \right),
\label{classicalM}
\end{eqnarray}
where $\theta_A$ and $\theta_B$ denote the arguments of $A$ and $B$. A
straightforward calculation gives
\begin{eqnarray}
\lefteqn{|\mathbf p|\left(|\mathbf{x}_E-\mathbf{x}_A|-|\mathbf{x}_E-\mathbf{x}_B|\right)}  \nonumber\\
& = &|\mathbf p|\left(\sqrt{l^2+\left(z+\frac{d}{2}\right)^2}-\sqrt{l^2+\left(z-\frac{d}{2}\right)^2}\right)\nonumber\\
& = & |\mathbf p|\frac{2zd}{\sqrt{l^2+\left(z+\frac{d}{2}\right)^2}+\sqrt{l^2+\left(z-\frac{d}{2}\right)^2}}\nonumber\\
& \approx & |\mathbf p|\frac{zd}{\sqrt{l^2+z^2}} \, ,
\label{classicald}
\end{eqnarray}
where we assumed that $d\ll l$. As a consequence, the maxima in the interference
pattern at the screen should be separated approximately by
\begin{equation}
    \frac{|\mathbf p| d}{2 \pi \hbar \sqrt{l^2+z^2}}.
\label{eq:IntMax}
\end{equation}
In principle, the separation of the maxima of interference fringes can depend
on the phase of the term $X_{x'\rightarrow (p'',x'')}$. A numerical calculation performed for
the  geometry of the interference system specified in Eqs.\ (\ref{eq:IntVec1}) and (\ref{eq:IntVec2})
and the range of parameters used in Fig.\ \ref{Fig:Int} shows that the combined term
$X_{x_A\rightarrow (p_E,x_E)}X^\star_{x_B\rightarrow (p_E,x_E)}$ in Eq.\ (\ref{eq:InterM})
remains close to a real number. Specifically, the ratio of the real part to the
imaginary part of $X_{x_A\rightarrow (p_E,x_E)}X^\star_{x_B\rightarrow (p_E,x_E)}$
 varies from the order of $10^2$ to $10^6$.
Consequently, the formula (\ref{eq:IntMax}) provides a satisfactory approximation.
Thus Eq.\ (\ref{eq:InterM}) can be written in the
approximate way as
\begin{eqnarray}
M & \approx & |X_{x_A\rightarrow (p_E,x_E)}|^2+|X_{x_B\rightarrow
(p_E,x_E)}|^2+ \nonumber  \\
&&\label{eq:IntApp2}
2|X_{x_A\rightarrow (p_E,x_E)}|\, |X_{x_B\rightarrow
(p_E,x_E)}|\cos\left(\frac{|\mathbf{p}_E|zd}{\hbar\sqrt{l^2+z^2}}\right).
\end{eqnarray}

\begin{figure}
\begin{center}
\includegraphics[width=0.8\textwidth]{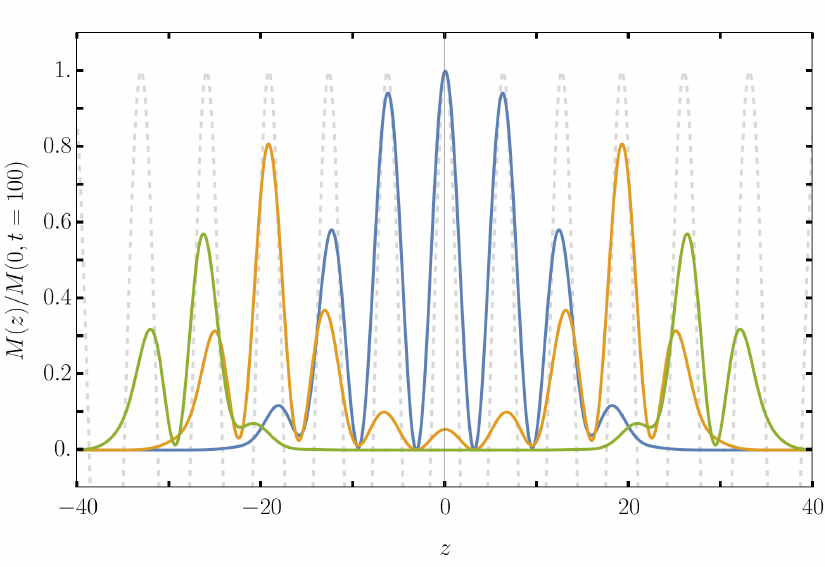}
\end{center}
\caption{
\label{Fig:Intm0}
Same as in Fig.\ \ref{Fig:Int}, but for $m = 0$ and  $M(z=0,t=100)\simeq 4.9 \times 10^{-5}$. We use analytic
formulas (\ref{m0final}).
The blue, orange, and green curves correspond to $t=100$, $t=102$, and $t=104$, respectively.
}
\end{figure}

The amplitudes $\mathcal{A}_{x_A\rightarrow (p_E,x_E)}$
and $\mathcal{A}_{x_B\rightarrow (p_E,x_E)}$ can also be calculated as follows:
\begin{eqnarray}
\mathcal A_{x' \to (p'',x'')} & \approx & \frac{1}{4 (\pi \hbar)^3 (\lambda_0 \lambda_3^3)^\frac{1}{4}} \exp \left( - i \frac{p'' x''}{2 \hbar} \right) \int_{\RNumb^3} d_0^3 p \left( \sqrt{m^2 + \mathbf p^2 + \epsilon} - \sqrt{m^2 + \mathbf p^2}  \right) \nonumber \\
&& \exp \left( - i \frac{\sqrt{m^2 + \mathbf p^2} \Delta x^0}{\hbar} \right) \exp \left( i \frac{\mathbf p \cdot \Delta \mathbf x}{\hbar} \right) \nonumber \\
&& \exp \left( - \frac{\left( \sqrt{m^2 + \mathbf p^2} + p_0'' \right)^2}{2 \hbar \lambda_0} \right) \exp \left( -\frac{(\mathbf p - \mathbf p'')^2}{2 \hbar \lambda_3} \right).
\end{eqnarray}
As before, the Taylor expansion with respect to $\epsilon$ gives, up to the
linear order,
\begin{eqnarray}
\mathcal A_{x' \to (p'',x'')} & \approx &
\frac{\epsilon}{8 (\pi \hbar)^3 (\lambda_0 \lambda_3^3)^\frac{1}{4}}
\exp \left( - i \frac{p'' x''}{2 \hbar} \right)
\int_{\RNumb^3}  \frac{d_0^3 p}{\sqrt{m^2 + \mathbf p^2}}
\exp \left( - i \frac{\sqrt{m^2 + \mathbf p^2} \Delta x^0}{\hbar} \right)
\nonumber \\
&& \exp \left( i \frac{\mathbf p \cdot \Delta \mathbf x}{\hbar} \right)
\exp \left( - \frac{\left( \sqrt{m^2 + \mathbf p^2} + p_0'' \right)^2}{2 \hbar
  \lambda_0} \right)
\exp \left( -\frac{(\mathbf p - \mathbf p'')^2}{2 \hbar \lambda_3} \right)\, .
\end{eqnarray}
Expressing $\mathbf p$ and $\mathbf p''$ in spherical coordinates
$(|\mathbf p|,\theta,\varphi)$ and $(|\mathbf p''|,\theta'',\varphi'')$,
respectively, one gets
\begin{eqnarray}
\mathbf p \cdot \Delta \mathbf x & = & |\mathbf p| |\Delta \mathbf x| \cos
\theta, \\
(\mathbf p - \mathbf p'')^2 & = & |\mathbf p|^2 + |\mathbf p''|^2 \nonumber \\
&& - 2 |\mathbf p||\mathbf p''| \left[ \cos \theta \cos \theta'' + \sin \theta
\sin \theta'' \cos (\varphi - \varphi'') \right].
\end{eqnarray}
Here the axis of the coordinate system is oriented along the vector $\Delta
\mathbf x$. Thus,
\begin{equation}
\int_0^{2 \pi} d \varphi
\exp\left( \frac{|\mathbf p||\mathbf p''| \sin \theta \sin\theta'' \cos(\varphi
- \varphi'')}{\hbar \lambda_3} \right) = 2 \pi I_0
\left( \frac{|\mathbf p||\mathbf p''| \sin \theta \sin \theta''}{\hbar
\lambda_3} \right),
\end{equation}
and consequently,
\begin{eqnarray}
\lefteqn{\mathcal A_{x' \to (p'',x'')}  = \frac{2 \pi \epsilon}{8 (\pi \hbar)^3 (\lambda_0 \lambda_3^3)^\frac{1}{4}} \exp \left( - i \frac{p'' x''}{2 \hbar} \right) } \nonumber \\
&& \int_0^\infty \frac{d|\mathbf p|}{\sqrt{m^2 + |\mathbf p|^2}} \int_0^\pi d \theta |\mathbf p|^2 \sin \theta \exp \left( - i \frac{\sqrt{m^2 + |\mathbf p|^2} \Delta x^0}{\hbar} \right) \nonumber \\
&& \exp \left( i \frac{|\mathbf p||\Delta \mathbf x| \cos \theta}{\hbar} \right) \exp \left( - \frac{\left( \sqrt{m^2 + |\mathbf p|^2} + p_0'' \right)^2}{2 \hbar \lambda_0} \right) \nonumber \\
&&  \exp \left( -\frac{|\mathbf p|^2 + |\mathbf p''|^2 - 2 |\mathbf p| |\mathbf p''| \cos \theta \cos \theta''}{2 \hbar \lambda_3} \right) I_0 \left( \frac{|\mathbf p||\mathbf p''| \sin \theta \sin \theta''}{\hbar \lambda_3} \right).
\end{eqnarray}
The integral with respect to $\theta$ can be computed analytically, assuming $\theta'' = 0$. One then has
\begin{eqnarray}
\lefteqn{\mathcal A_{x' \to (p'',x'')}  = \frac{2 \pi \epsilon}{8 (\pi \hbar)^3 (\lambda_0 \lambda_3^3)^\frac{1}{4}} \exp \left( - i \frac{p'' x''}{2 \hbar} \right)  \exp \left( - \frac{m^2 + (p_0'')^2}{2 \hbar \lambda_0} \right) }  \\
&& \int_0^\infty \frac{d|\mathbf p|}{\sqrt{m^2 + |\mathbf p|^2}} \frac{\hbar |\mathbf p| \lambda_3}{ |\mathbf p''| + i |\Delta \mathbf x| \lambda_3} \left[ - 1 + \exp \left( \frac{2|\mathbf p| (|\mathbf p''| + i |\Delta \mathbf x| \lambda_3)}{\hbar \lambda_3} \right) \right] \nonumber \\
&& \exp \left( - \frac{(|\mathbf p| + |\mathbf p''|)^2}{2 \hbar \lambda_3} - \frac{\sqrt{m^2 + |\mathbf p|^2} p_0''}{\hbar \lambda_0} - \frac{i \sqrt{m^2 + |\mathbf p|^2} \Delta x^0}{\hbar} - \frac{|\mathbf p|^2}{2 \hbar \lambda_0} - i \frac{|\mathbf p| |\Delta \mathbf x|}{\hbar} \right). \nonumber
\end{eqnarray}
The remaining integral with respect to $|\mathbf p|$ can be computed
analytically in the limit of $m \to 0$, leading to the following result:
\begin{eqnarray}
\mathcal A_{x' \to (p'',x'')}  & = & \frac{2 \pi \epsilon}{8 (\pi \hbar)^3 (\lambda_0 \lambda_3^3)^\frac{1}{4}} \exp \left( - i \frac{p'' x''}{2 \hbar} \right)  \exp \left( - \frac{(p_0'')^2}{2 \hbar \lambda_0} \right) \frac{\hbar \lambda_3}{ |\mathbf p''| + i |\Delta \mathbf x| \lambda_3} \nonumber \\
&& \left\{- \exp Y_+ \int_0^\infty d|\mathbf p| \exp \left[ - A (|\mathbf p| + X_+)^2 \right] \right. \nonumber \\
&& \left. + \exp Y_-  \int_0^\infty d|\mathbf p| \exp \left[ - A (|\mathbf p| + X_-)^2 \right] \right\} \nonumber \\
& = & \frac{2 \pi \epsilon}{8 (\pi \hbar)^3 (\lambda_0 \lambda_3^3)^\frac{1}{4}} \exp \left( - i \frac{p'' x''}{2 \hbar} \right)  \exp \left[ - \frac{(p_0'')^2}{2 \hbar \lambda_0} \right] \frac{\hbar \lambda_3}{ |\mathbf p''| + i |\Delta \mathbf x| \lambda_3} \nonumber \\
& &  \frac{\sqrt{\pi}}{2 \sqrt{A}} \left[ \exp Y_+ \mathrm{erfc} \, \left( \sqrt{A} X_+ \right) - \exp Y_- \mathrm{erfc} \, \left( \sqrt{A} X_- \right) \right],
\label{m0final}
\end{eqnarray}
where
\begin{eqnarray}
A   & = & \frac{1}{2 \hbar} \left( \frac{1}{\lambda_0} + \frac{1}{\lambda_3}
\right),  \\
X_\pm & = & \frac{\pm|\mathbf{p}''| \lambda_0 +\left[ p_0'' \pm i \left( |\Delta
\mathbf{x}| \pm \Delta x^0 \right) \right]\lambda_3}{\lambda_0 + \lambda_3} ,
\\
Y_\pm & = & \frac{1}{2 \hbar \lambda_0 (\lambda_0 + \lambda_3)} \left\{\pm
|\mathbf{p}''| \lambda_0 \left[ 2p_0'' \mp |\mathbf{p}''| \pm 2 i \left( |\Delta \mathbf{x}| \pm \Delta x^0 \right)\lambda_0   \right]  \right. \nonumber \\
&& \left.  + \left[ p_0'' \pm  i \left( |\Delta \mathbf{x}| \pm \Delta x^0 \right)\lambda_0\right]^2 \lambda_3 \right\},
\end{eqnarray}
and $\mathrm{erfc}$ denotes the error function given by
\begin{equation}
\label{erfc}
    \mathrm{erfc}(z) = \frac{2}{\sqrt{\pi}} \int^\infty_z e^{-t^2} dt.
\end{equation}

Figure \ref{Fig:Int} illustrates an interference pattern defined by formula
(\ref{eq:InterM}), as a function of position $z$ on a screen.  The geometrical
configuration of the interference system is represented by the system of vectors
defined by (\ref{eq:IntVec1}) and (\ref{eq:IntVec2}). The graph shown in
Fig.~\ref{Fig:Int} has been normalized by the value of the amplitude $M$
corresponding to the time $t=100$ and the position on the screen
$z=0$. Different colors correspond to different times of the measurement. The
blue curve is computed for $t=100$, the orange one for $t=102$, and the green
one for $t=104$. The gray dashed line represents the frequency of interference
fringes predicted by the formula (\ref{eq:IntApp2}).

For greater times, the largest interference fringes become more distant from
each other. For example, this occurs at times $t = 102$ and $t = 104$,
respectively.  This behavior can be explained by the shape of the probability
amplitude for the test particle, which takes on a croissant-like form, as
illustrated in Fig. \ref{Fig:P-stwa10}, for $3\lambda_3 = 30$. This specific
type of shape for the probability amplitude also manifests itself for smaller
$\lambda_0$ and $\lambda_3$ and for a sufficiently large time. The interference
fringes for an early time of measurement are generated by the central part of
the amplitude, whereas for subsequent times, interference fringes are formed by
the tails of transition amplitudes.

%%%%%%%%%%%%%%%%%%%%%%%%%%%%%%%%%%%%%%%%%%%%
\section{Summary and conclusions}\label{Con}
%%%%%%%%%%%%%%%%%%%%%%%%%%%%%%%%%%%%%%%%%%%%

In this paper we used the Heisenberg-Weyl group to construct the
quantum configuration space, consisting of coherent states. The Heisenberg-Weyl group reproduces standard canonical commutation
relations among positions and momenta. In addition, its natural parametrization
happens to be compatible with the Minkowski spacetime. The carrier space of that representation is used as the Hilbert space of the
considered quantum system. Since the representation is irreducible, there
exists the decomposition of the unity in the carrier space that can be used for
mapping of almost any classical observable onto a quantum observable in the
considered Hilbert space.

Quantum evolution of a test particle is constructed by making use of the
generalized eigenstates of the quantum Hamiltonian, associated with the classical motion of the particle. More precisely,
the generalized eigenvalues of that Hamiltonian satisfy a relation
similar to the one defining causal geodesics corresponding to the classical motion of test particles. The
projection operator used to define the transition amplitudes of the test
particle is constrained by the orthochronous part of the Lorentz group.

We have applied our model to the calculation of the transition
probabilities of a quantum particle from one state to another.  The
results indicate that the probability is concentrated along the classical
geodesic with some smearing. To some extent, this results from a suitable
choice of the so-called fiducial vector, which is used to generate all coherent
states. The amplitude of the probability distribution becomes smaller and more dispersed
as the evolution time increases.

The transition amplitude is invariant with respect to translations in the spacetime
for any form of the fiducial vector. Its invariance with respect to Lorentz
transformations requires special choice of the fiducial vector which remains an open problem
to be addressed elsewhere.

Another application of our formalism concerns a random walk of a test
particle. In this case, the quantum transition probability distribution is used
to calculate subsequent steps of this stochastic process. The random walk
trajectories are represented by polylines with vertices at the particle
positions in subsequent steps. Quantum particle positions stay close to the
classical path, but depart from it more and more as the evolution time
increases.  This is consistent with the smearing of the time dependence of the
probability distribution.

We have also examined the interference phenomenon. The transition amplitudes are
defined for the initial and final states with a set of intermediate states. Our
approach seems to be consistent with Feynman's idea of a quantum propagation
\cite{Feynman}. The considered double-slit interference yields expected
interference patterns both for particles with positive and zero rest mass
(photons).

Our semiclassical construction is designed keeping in mind a future
generalization to curved spacetimes. Certainly, the simple setup considered in
this paper cannot be applied to the diffraction on astrophysical objects like
black holes, where the interference occurs along (closed) curves with
``interference slots'' distributed continuously. On the other hand our results
indicate that the integral quantization formalism is open for such issues.
We believe that the ideas presented in our article can be extended to other
spacetimes.

%%%%%%%%%%%%%%%%%%%%%%%%%%%%%%%%%%%%%%%%%%%%%%%%%%%%%%%%%%%%%%%%%%%%%%%%%%%%%%%%%%%%%%%%%%%%%%%%%%%%%%%%%
\acknowledgments
%\tolerance=5000
\noindent A.C. \ acknowledges support from the Austrian Science Fund (FWF) 10.55776/PAT7614324.

%%%%%%%%%%%%%%%%%%%%%%%%%%%%%%%%%%%%%%%%%%%%%%%%%%%%%%%%%%%%%%%%%%%%%%%%%%%%%

\appendix
%%%%%%%%%%%%%%%%%%%%%%%%%%%%%%%%%%%%%%%%%%%%%%%%%%%%%%%%%

\section{Overlaps in the configuration space}\label{Prob}
%%%%%%%%%%%%%%%%%%%%%%%%%%%%%%%%%%%%%
In this Appendix, we compute the transition probability between coherent states (their overlap). Assuming a fiducial vector given by Eq.\ (\ref{Ham17}), one obtains
\begin{eqnarray}
\lefteqn{\BraKet{p'',x''}{p',x'}} \nonumber\\
&=&\int_{\RNumb^4}d^4\xi
\left[
\exp\left(-\frac{ip''_\nu x''^\nu}{2\hbar}\right)\exp\left(-\frac{i p''_\nu \xi^\nu}{\hbar}\right)
\prod_{\mu=0}^3
\left(\frac{\lambda_\mu}{\pi\hbar}\right)^{\frac{1}{4}}
\exp\left(-\frac{\lambda_\mu (\xi^\mu-x''^\mu)^2}{2\hbar}\right)
\right]^\star \nonumber \\
&& \left[
\exp\left(-\frac{ip'_\nu x'^\nu}{2\hbar}\right)\exp\left(-\frac{i p'_\nu \xi^\nu}{\hbar}\right)
\prod_{\mu=0}^3
\left(\frac{\lambda_\mu}{\pi\hbar}\right)^{\frac{1}{4}}
\exp\left(-\frac{\lambda_\mu (\xi^\mu-x'^\mu)^2}{2\hbar}\right)
\right] \nonumber \\
&=&\exp\left(i\frac{p''_\nu x''^\nu-p'_\nu x'^\nu}{2\hbar}\right)
\prod_{\mu=0}^3
\sqrt{\frac{\lambda_\mu}{\pi\hbar}}
\exp\left(-\frac{\lambda_\mu((x''^\mu)^2+(x'^\mu)^2)}{2\hbar}\right) \nonumber \\
&& \int_\RNumb d\xi^\mu \exp\left(\frac{i\xi^\mu(p''_\mu-p'_\mu)}{\hbar}-\frac{\lambda_\mu((\xi^\mu)^2-\xi^\mu(x''^\mu+x'^\mu))}{\hbar}\right) \nonumber \\
&=& \exp\left(i\frac{p''_\nu x''^\nu-p'_\nu x'^\nu}{\hbar}\right)\exp\left(i\frac{p''_\nu x'^\nu-p'_\nu x''^\nu}{2\hbar}\right) \nonumber \\
&&\prod_{\mu=0}^3
\exp\left(-\frac{\lambda_\mu(x''^\mu-x'^\mu)^2+\frac{1}{\lambda_\mu}(p''_\mu-p'_\mu)^2}{4\hbar}\right).
\end{eqnarray}
Hence the transition probability reads
\begin{equation}
\label{cohbraket}
|\BraKet{p'',x''}{p',x'}  |^2  = \prod_{\mu=0}^3
\exp\left(-\frac{\lambda_\mu(x''^\mu-x'^\mu)^2
+\frac{1}{\lambda_\mu}(p''_\mu-p'_\mu)^2}{2\hbar}\right)
\, .
\end{equation}
In this case the overlap between two coherent states is given by a four-dimensional Gaussian function.

\section{Transition probability of a massless test particle}
\label{App-massless}
%%%%%%%%%%%%%%%%%%%%%%%%%%%%%%%%%%%%%

For a massless test particle, the mass layer \eqref{eq:MassLayer} has the
following form:
\begin{equation}
\label{eq:MasslessLayer}
\mathcal{J}_{0,\epsilon}
= \left\{ p \colon  -\sqrt{\mathbf{p}^2+\epsilon} \leq p_0
\leq -\sqrt{\mathbf{p}^2}, \,\, \mathbf{p} \in \RNumb^3  \right\} \,.
\end{equation}
Repeating the calculations discussed in Sec.\ \ref{sec:Transition_prob}, one obtains
\begin{eqnarray}
\mathcal{A}_{0,\epsilon} & = & \Bra{p'',x''} P_{\mathcal{J}_{0,\epsilon}} \Ket{p',x'} \nonumber \\
& \approx & \epsilon \, \pi \exp\left(i \frac{p'\,x'-p''\,x''}{2\hbar} \right)\, \mathcal{B}
 \int_0^\infty d |\mathbf p|  \exp\left( -\frac{1}{\hbar \lambda_0} [|\mathbf p| + \bar{p}_0]^2 \right)  \nonumber \\
&&  \exp\left(\frac{i}{\hbar} \left[ -|\mathbf p| (x''^{0}-x'^{0}) \right] \right) \exp\left(
-\frac{1}{\hbar \lambda_3} \left[ |\mathbf p|^2 + |\bar{\mathbf p}|^2 \right] \right) \frac{2}{\left( \frac{i}{\hbar} |\boldsymbol \zeta| + \frac{2}{\hbar \lambda_3} |\bar{\mathbf p}|  \right)} \nonumber \\
&&  \sinh \left( \frac{i}{\hbar}  |\mathbf p| |\boldsymbol \zeta| + \frac{2}{\hbar \lambda_3} |\mathbf p| |\bar{\mathbf p}|  \right).
\label{eq:AverP0TransAmpSphColinarMassless}
\end{eqnarray}
Expression (\ref{eq:AverP0TransAmpSphColinarMassless}) corresponds directly to the limit
\begin{equation}
    \mathcal{A}_{0,\epsilon}  = \lim_{m\to 0} \mathcal{A}_{m,\epsilon}
    \end{equation}
of Eq.\ \eqref{eq:AverP0TransAmpSphColinar}.
The integral appearing in Eq.\ (\ref{eq:AverP0TransAmpSphColinarMassless}) can be calculated analytically. Rearranging terms in (\ref{eq:AverP0TransAmpSphColinarMassless}), we get
\begin{eqnarray}
\mathcal{A}_{0,\epsilon} & \approx &  \frac{2\epsilon \, \pi }{\left( \frac{i}{\hbar} |\boldsymbol \zeta| + \frac{2}{\hbar \lambda_3} |\bar{\mathbf p}|  \right)} \exp \left(i \frac{p'\,x'-p''\,x''}{2\hbar}\right)\, \mathcal{B}
 \int_0^\infty d |\mathbf p|   \exp\left(-\frac{1}{\hbar}\left[ \frac{1}{\lambda_0} + \frac{1}{\lambda_3} \right] |\mathbf p|^2 \right)  \nonumber \\
&&  \exp\left(-\frac{1}{\hbar} \left[ i (x''^{0}-x'^{0}) + \frac{2}{\lambda_0} \bar{p}_0 \right] |\mathbf p| \right)
\exp\left(-\frac{1}{\hbar} \left[ \frac{\bar{p}_0^2}{\lambda_0} + \frac{|\bar{\mathbf p}|^2}{\lambda_3} \right] \right) \nonumber \\
&&  \sinh \left[ |\mathbf p|\left( \frac{i}{\hbar}  |\boldsymbol \zeta| + \frac{2}{\hbar \lambda_3}  |\bar{\mathbf p}| \right)  \right].
\end{eqnarray}
Denoting
\begin{eqnarray}
    \alpha &=& \frac{1}{\hbar}\left( \frac{1}{\lambda_0} + \frac{1}{\lambda_3} \right),  \quad  \beta =  -\frac{1}{\hbar} \left[ i (x''^{0}-x'^{0}) + \frac{2}{\lambda_0} \bar{p}_0 \right] , \nonumber \\
    \gamma &=& -\frac{1}{\hbar} \left( \frac{\bar{p}_0^2}{\lambda_0}+ \frac{|\bar{\mathbf p}|^2}{\lambda_3} \right) ,\quad \omega = \frac{1}{\hbar}  \left( i|\boldsymbol \zeta| + \frac{2}{ \lambda_3}  |\bar{\mathbf p}| \right),  \nonumber
\end{eqnarray}
one can write
\begin{eqnarray}
    && \int_0^\infty d|\mathbf p| \; e^{ -\alpha |\mathbf p|^2 + \beta |\mathbf p| + \gamma } \sinh \left( |\mathbf p| \omega \right) = \nonumber \\
    && \frac{1}{2} \int_0^\infty d|\mathbf p| \; e^{ -\alpha |\mathbf p|^2 + (\beta + \omega) |\mathbf p| + \gamma }  - \frac{1}{2} \int_0^\infty d|\mathbf p| \; e^{ -\alpha |\mathbf p|^2 + (\beta - \omega) |\mathbf p| + \gamma } \nonumber \\
    && = \frac{1}{2} e^{\frac{(\beta + \omega)^2}{4\alpha} + \gamma} \int_0^\infty d|\mathbf p| \; e^{-\alpha\left( |\mathbf p|+\frac{\beta - \omega}{2\alpha} \right)^2} -  \frac{1}{2} e^{\frac{(\beta - \omega)^2}{4\alpha} + \gamma} \int_0^\infty d |\mathbf p| \; e^{-\alpha\left(|\mathbf p| - \frac{\beta - \omega}{2\alpha} \right)^2} \nonumber \\
    && =  \frac{1}{\sqrt{2\alpha}} e^{\frac{(\beta + \omega)^2}{4\alpha} + \gamma} \int^{\frac{\beta + \omega}{\sqrt{2\alpha}}}_{-\infty} du\; e^{- u^2/2} - \frac{1}{\sqrt{2\alpha}}e^{\frac{(\beta - \omega)^2}{4\alpha} + \gamma} \int^{\frac{\beta - \omega}{\sqrt{2\alpha}}}_{-\infty} d\tilde{u}\; e^{-\tilde{u}^2/2 },
\end{eqnarray}
where in the last step, we have substituted  $u = - \sqrt{2\alpha} \left(|\mathbf p|-\frac{\beta + \omega}{2\alpha} \right) $ and $\tilde{u} = - \sqrt{2\alpha} \left( |\mathbf p| - \frac{\beta - \omega}{2\alpha}\right)$.
This yields
\begin{eqnarray}
    \mathcal{A}_{0,\epsilon} & \approx &  \frac{\pi^{3/2} \hbar \epsilon }{\sqrt{\alpha}\left[ i |\boldsymbol \zeta| + \frac{2}{ \lambda_3} |\bar{\mathbf p}|  \right]}\exp\left(i \frac{p'\,x'-p''\,x''}{2\hbar}\right)\, \mathcal{B}  e^\gamma \nonumber \\
    && \left\{ e^{\frac{(\beta + \omega)^2}{4\alpha} } \left[1+\mathrm{erf}\left( \frac{\beta + \omega}{2\sqrt{\alpha}} \right) \right] -
    e^{\frac{(\beta - \omega)^2}{4\alpha}} \left[1+ \mathrm{erf}\left( \frac{\beta - \omega}{2\sqrt{\alpha}} \right)
    \right]\right\},
\end{eqnarray}
where
\begin{equation}
    \mathrm{erf}(z) = \frac{2}{\sqrt{\pi}}\int^z_0 e^{-t^2}dt.
\end{equation}
Finally
\begin{eqnarray}
    \mathcal{A}_{0,\epsilon} & \approx & \frac{  (\pi \hbar)^{3/2}  \epsilon}{\left[ i |\boldsymbol \zeta| + \frac{2}{ \lambda_3} |\bar{\mathbf p}|  \right] \sqrt{\frac{1}{\lambda_0} + \frac{1}{\lambda_3}}}  \exp\left(i \frac{p'\,x'-p''\,x''}{2\hbar} -\frac{1}{\hbar} \left[ \frac{\bar{p}_0^2}{\lambda_0}+ \frac{|\bar{\mathbf p}|^2}{\lambda_3} \right]\right)\, \mathcal{B} \nonumber \\
    && \left\{ \exp\left( \frac{Z_-^2}{4 \hbar \left( \frac{1}{\lambda_0} +  \frac{1}{\lambda_3} \right)} \right) \left[ 1 + \mathrm{erf}\left(  \frac{Z_-}{ 2\sqrt{\hbar \left( \frac{1}{\lambda_0} - \frac{1}{\lambda_3} \right)}}  \right) \right]\right. \nonumber \\
    && \left. - \exp\left( \frac{Z_+^2}{4 \hbar \left( \frac{1}{\lambda_0} + \frac{1}{\lambda_3} \right)} \right)
     \left[ 1 + \mathrm{erf}\left(  \frac{Z_+}{ 2\sqrt{\hbar \left( \frac{1}{\lambda_0} - \frac{1}{\lambda_3} \right)}}  \right) \right]
    \right\}
\end{eqnarray}
where $\Delta x^0 =x''^{0} - x'^{0}$ and
\begin{equation}
Z_\pm = -i \left(\Delta x^0 \pm |\boldsymbol \zeta| \right) -  2 \left( \frac{\bar{p}_0}{\lambda_0} \pm \frac{|\bar{\mathbf p}|}{\lambda_3}  \right).
\end{equation}

Plots of $|\mathcal{A}_{0,\epsilon}(\mathbf{x}'')/\epsilon|^2$ can be prepared in a straightforward way. They are qualitatively similar to Figs.\ \ref{Fig:P-stwa01} and \ref{Fig:P-stwa10}.

%%%%%%%%%%%%%%%%%%%%%%%%%%%%%%%%%%%%%%%%%%%%%%%%%%%%%%%%%%%%%%%%%%%%%%%%%%
%%%%%%%%%%%%%%%%%%%%%%%%%%%%%%%%%%%%%%%%%%%%%%%%%%%%%%%%%%%%%%%%%%%%%%%%%%%

\end{document}